\documentclass[sigconf]{acmart}
\usepackage{multirow, tabularx, ragged2e,subfig}

\AtBeginDocument{%
  }

\copyrightyear{2026}
\acmYear{2026}
\setcopyright{cc}
\setcctype{by-nc-nd}
\acmConference[DIS '26]{Designing Interactive Systems Conference}{June 13--17, 2026}{Singapore, Singapore}
\acmBooktitle{Designing Interactive Systems Conference (DIS '26), June 13--17, 2026, Singapore, Singapore}
\acmDOI{10.1145/3800645.3812938}
\acmISBN{979-8-4007-2563-0/2026/06}

\begin{document}


\title[Towards Real-World Validity in Generative AI Benchmarks]{Towards Real-World Validity in Generative AI Benchmarks: Understanding and Designing Domain-Centered Evaluations for Journalism Practitioners}

\author{Charlotte Li}
\email{charlotte.li@u.northwestern.edu}
\orcid{0009-0004-8914-3717}
\affiliation{%
  \institution{Northwestern University}
  \city{Evanston}
  \state{Illinois}
  \country{USA}
}

\author{Nick Hagar}
\email{nhagar@u.northwestern.edu}
\orcid{0000-0001-5110-3737}
\affiliation{%
  \institution{Northwestern University}
  \city{Evanston}
  \state{Illinois}
  \country{USA}
}

\author{Sachita Nishal}
\email{nishal@u.northwestern.edu}
\orcid{0000-0001-6192-6091}
\affiliation{%
  \institution{Northwestern University}
  \city{Evanston}
  \state{Illinois}
  \country{USA}
}

\author{Jeremy Gilbert}
\email{jeremy.gilbert@northwestern.edu }
\orcid{0009-0008-2585-0972}
\affiliation{%
  \institution{Northwestern University}
  \city{Evanston}
  \state{Illinois}
  \country{USA}
}

\author{Nicholas Diakopoulos}
\email{nad@northwestern.edu}
\affiliation{%
  \institution{Northwestern University}
  \city{Evanston}
  \state{Illinois}
  \country{USA}
}

\renewcommand{\shortauthors}{Li et al.}

\begin{abstract}
Benchmarks play a significant role in how technology companies communicate about model capabilities and how researchers and the public understand generative AI systems. However, existing benchmarks have been criticized for their failure to adequately capture real-world usages (i.e. ecological validity) or to measure underlying concepts (i.e. construct validity). Building on approaches in HCI, we adopt a human-centered design process to address such critiques. Working within the journalism domain we engaged 23 professionals in a workshop which informed the design of a domain-oriented evaluation ``cookbook''. Our workshop findings surface domain-specific challenges and tensions faced by designers in translating specific tasks to evaluation constructs, aligning metrics with domain-specific values, and balancing needs among different stakeholders when constructing evaluations. Through an instantiation of design-based approaches for benchmark creation in the journalism domain, this work not only produces an evaluation structure for journalism practitioners to experiment with, but also lays out design requirements for AI evaluations that are contextualized, value-aligned, and cultivate evaluative literacy for domain end-users.

\end{abstract}



\received[accepted]{25 April 2026}

\maketitle

\section{Introduction}
Opening the technology column of any news site around the world today, a reader would not be surprised to find yet another article about the cutting-edge capabilities of some newly released language model. The development and release of frontier models, by technology companies around the world \cite{singh2025openaigpt5card, deepseekai2025deepseekv32pushingfrontieropen} or on open-source platforms, has proliferated in the public discourse. One of the features readers can expect in these releases is tables and graphs of ``performance benchmarks,'' boasting of each new model's state of the art capabilities \cite{anthropic2025claude4, openai2025gpt5}. These benchmarks, which are datasets of tasks and metrics for evaluating AI models, stem from traditions in the machine learning community to provide a standardized comparison across different models in order to track model development progress and foster a collaborative community of practice \cite{raji2021ai}. However, in the current era of commercial generative models, benchmarks are also used as a competitive battleground for commercial models and advertisements of model capabilities in order to expand the market of users \cite{10.1145/3630106.3659012}. This change in benchmark utility implies an expansion of benchmark users from model developers with extensive knowledge in machine learning to a broad group of people with expertise in a diverse set of fields beyond machine learning.

With changes in the desiderata for benchmarks and evaluations of generative models and systems, several criticisms have surfaced regarding the design of existing benchmarks, especially about whether measurements are representative of how models are used in the wild (i.e., ecological validity) and whether benchmarks test what they claim to test (i.e., construct validity) \cite{saxon2024benchmarksmicroscopesmodelmetrology, liao2025rethinkingmodelevaluationnarrowing, ethayarajh2021utilityeyeusercritique}. These criticisms invite sociotechnical approaches for understanding and evaluating the effectiveness and impact of new generative models in different usage scenarios across various domains and industries \cite{reuel2024betterbenchassessingaibenchmarks, patwardhan2025gdpvalevaluatingaimodel}. Specifically, researchers have begun to employ human-centered and design-based methods for creating evaluation approaches that address issues of validity. For example, Liu et al. \cite{liu2024ecbdevidencecenteredbenchmarkdesign} applied Evidence-Centered Design (ECD) to reinforce construct validity of measurements included in benchmarks. Separately, Liao and Xiao \cite{liao2025rethinkingmodelevaluationnarrowing} argued that an understanding and formalization of downstream applications of generative models is necessary for ecologically valid evaluations. 

We expand this emerging line of research on more human-centered evaluation approaches for generative AI by focusing on one specific domain: \textit{journalism}, such as is reflected in professional news production practices. Journalism makes for a suitable domain of study because it has seen rapid uptake of generative AI across a range of use cases \cite{nishal2024envisioning-891, cools2024uses-635, D’haeseleer13082025, 10.13140/RG.2.2.31540.05765} that can vary by context \cite{10.63744@svxDtDD45mvw}, and it has well-established commitments to specific domain values among practitioners that may come into tension with existing language models and tools \cite{10.1145/3715336.3735717}. This creates conditions where practitioners are increasingly familiar with generative AI but are also still uncertain of its performance and alignment with their specific values and use cases. While not the only possible domain where one could advance such research (e.g. medicine, law, or other professional practices may also be suitable), we are able to leverage our own domain knowledge and contacts to study this field in particular.

In this work we engage journalism practitioners in a workshop that serves to center their practices, values, and contexts of work in shaping AI evaluations for the domain \cite{10.1145/3630106.3658992}. Our work addresses two research questions oriented around understanding benchmarking needs in this specific domain, but also with an eye towards how this process might inform benchmark design more broadly. More specifically we are guided by the following questions: (1) \textit{What are the needs and challenges surrounding the design of a domain-oriented generative model benchmark?} and (2) \textit{How might practitioner-centered design processes address existing criticisms about validity in benchmark design?} To explore these research questions, we conducted an observational study at a day-long workshop with 23 journalism practitioners from various newsrooms in the U.S. who occupy newsroom roles from reporter and technologist to strategy and management, noting discussions surrounding evaluation of generative AI models in newswork across various use cases and with respect to domain values. This specific study set-up was chosen to represent various stakeholders in the domain that make decisions surrounding language model uses, while an observational approach allowed for interactions and conversations to occur naturally among participants. We qualitatively analyzed observations and notes from the workshop, detailing needs, tensions, and challenges for creating context-specific model evaluations for journalism practitioners. Motivated by a workshop discussion that touched on the utility of evaluation templates as opposed to a canonical domain benchmark, we further instantiate these design challenges through the design of a journalism evaluation cookbook prototype for an information extraction use case.

Our findings from the workshop and reflections from the prototype design process help articulate a set of design challenges surrounding constructing model evaluations that capture news practitioners' real-world use and encode domain-specific values. We discuss implications of this work for empowering practitioners to create evaluations suited for their own use cases and partake in benchmarking practices, applying design-based methods to creating generative model benchmarks to ground evaluations in downstream use and increase validity, and considering domain-level real world constraints when designing with domain practitioners. The contributions of this research are twofold. Firstly, this research identifies design challenges for creating journalism-specific evaluation systems, which extends a line of research on domain-specific measurements of generative AI systems to the domain of journalism. These challenges point to a gap between existing benchmark evaluations and the desiderata of practitioner-centered evaluation, and they invite further sociotechnical approaches that focus on literacy and empowerment of news practitioners. Moreover, this research advances a line of research on human-centered design-based approaches to address validity criticisms of existing generative model benchmarks through engaging domain practitioners in the process of benchmark design. In turn, this research shines a light on several areas in participatory generative model construction and evaluation, such as data access and value alignment, where design-based methods could be of further utility.

\section{Related Work}
In this section, we first provide an overview of common critiques of existing benchmarks, especially those of ecological validity in popular benchmarks, and research that calls for the incorporation of downstream use cases to address ecological validity issues. Then we provide an overview of previous work on model evaluations that incorporate real-world contexts, either through scenario-building or human-centered methods, and we situate our work within this literature.

\subsection{The Challenges of Benchmarking}
Benchmarks are measurement instruments consisting of a dataset or a set of datasets and a metric, characterizing tasks or abilities of machine learning models. They are adopted by communities of researchers as a standardized measurement of models’ abilities to accomplish certain tasks and provide comparison across models \cite{raji2021ai}. Specifically, since early AI developments focused on creating models that specialized in certain tasks, such as optical character recognition (OCR), benchmarks and leaderboards were designed to standardize the measure of these specialized tasks and promote development of ``better'' performing models among the machine learning community \cite{10.1145/3630106.3659012}. This benchmarking practice for specific tasks continued as more models were developed, even for recently published general purpose large language models: measuring systems’ abilities to generate code \cite{matton2024leakagecodegenerationevaluation, chen2021evaluatinglargelanguagemodels,austin2021programsynthesislargelanguage, yu-etal-2018-spider}, reason based on common sense \cite{Bisk2019PIQARA, zellers-etal-2019-hellaswag, sakaguchi2019winograndeadversarialwinogradschema}, comprehend and use texts \cite{dua2019dropreadingcomprehensionbenchmark, rajpurkar-etal-2018-know, lai-etal-2017-race}, perform arithmetic operations \cite{cobbe2021trainingverifierssolvemath,hendrycks2021measuring}, and so on. In addition to benchmarks that measure a specialized task, recent efforts also focused on creating benchmarks that measure the general intelligence of models \cite{liang2023holisticevaluationlanguagemodels, srivastava2023imitationgamequantifyingextrapolating, zhong2023agievalhumancentricbenchmarkevaluating}. Other benchmarking approaches have focused on soliciting crowd-sourced user evaluations of model performance by creating platforms where users can actively participate in pair-wise ratings of different model outputs, such as Chatbot Arena \cite{chiang2024chatbotarenaopenplatform} (now just https://arena.ai/) and GenAI Arena \cite{jiang2024genaiarenaopenevaluation}.

While benchmarks play a vital role in defining the standards that newly developed AI systems strive for and are persistently used when marketing new models’ capabilities, there is a growing chorus of criticism regarding the effectiveness and validity of existing benchmarks. These criticisms, often drawing on literature reviews and qualitative studies, speak to the quality of data included in benchmarks, e.g. incorrect answers \cite{10.1145/3615355} and contamination which effectively allow the model to “cheat” on a benchmark by having the answers in its training data \cite{reuel2024betterbenchassessingaibenchmarks}, the ability to account for user perspectives and downstream use cases, i.e. ecological validity \cite{saxon2024benchmarksmicroscopesmodelmetrology, liao2025rethinkingmodelevaluationnarrowing, ethayarajh2021utilityeyeusercritique, 10.1145/3708359.3712152}, the ability to generalize to concepts benchmarks aim to test, i.e. construct validity \cite{liu2024ecbdevidencecenteredbenchmarkdesign, saxon2024benchmarksmicroscopesmodelmetrology, 10.1145/3615355}, reproducibility of benchmark evaluations \cite{biderman2024lessonstrenchesreproducibleevaluation}, and durability, including saturation and maintenance of benchmarks \cite{saxon2024benchmarksmicroscopesmodelmetrology}. Even though the current state of global model benchmarking addresses some of these concerns regarding user perspectives and contamination, the lack of context in test cases presented to users for rating still results in less ecologically valid and reproducible benchmarks.

Given concerns surrounding validity of existing benchmarks and the expansion of downstream usage of generative AI tools, research has taken on a more critical lens toward understanding the utility of benchmarks and designing benchmarks that suit downstream use cases in various domains. In one example, Hardy et al. \cite{10.1145/3708359.3712152} focused on exploring the day-to-day utility of benchmarks with practitioners from tech, academia, and policy through an interview study and recommended that benchmarks be situated in real-world contexts and incorporate domain expertise in order to expand downstream utility and improve ecological validity of benchmarks. Similarly, Liao and Xiao \cite{liao2025rethinkingmodelevaluationnarrowing} argues model evaluations should move beyond metrics and adapt to sociotechnical requirements to capture real-world contexts of how models are used, while also laying out pragmatic challenges and trade-offs, such as cost, in conducting evaluations that capture necessary contexts and human requirements.

Our study extends this line of research by developing an understanding of the utility of model evaluations for the journalism domain, where adoption of generative AI tools and debates about them are happening simultaneously \cite{10.13140/RG.2.2.31540.05765, cools2024uses-635}. For practitioners in journalism, while some existing benchmarks such as Factcheck-Bench \cite{wang2024factcheckbenchfinegrainedevaluationbenchmark} and IssueBench \cite{rottger2025issuebenchmillionsrealisticprompts} partially address specific tasks or issues in their practice like fact-checking and media bias, they fail to capture the full extent of their work and how different context and values could come into play at various stages. Prior work in evaluating LLMs for editorial use also generated NewsBench \cite{li2024newsbench-770}, which includes expert written test questions for writing proficiency and safety adherence, but might be limited to values that exist in the Chinese journalism ecosystem. An industry-driven evaluation of LLM-based AI assistants \cite{bbcebu2025newsintegrity} focused on issues from an end-user perspective (i.e. answering news questions) rather than from a newsroom production standpoint. Regardless, to our knowledge, no study has examined the use of language model benchmarks in practice by journalism practitioners. Hence, through engaging journalism practitioners in designing evaluations that suit their practices, our study gathers criteria for situating model evaluations in real-world contexts and in turn addressing criticisms of ecological validity in existing benchmarking practices.

\subsection{Evaluations Rooted in Real-World Requirements}
Researchers have taken various different approaches to capturing realism in model evaluation constructs. Kuo et al. \cite{10.1145/3613904.3642278} developed a community-based data curation tool for creating evaluation datasets from Wikipedia for assessing AI-based content moderation tools deployed in the community, allowing community members to select data points and input personal judgments into a benchmark. Other researchers have experimented with usage-scenario-based evaluation when there is no centralized community that can generate data curation and evaluation insights. For example, Liang et al. \cite{liang2023holisticevaluationlanguagemodels} took the approach of creating a series of scenarios by crossing common natural language generation tasks with domain contexts (who, what, when) coupled with different languages for evaluating generative models in different contexts. Similarly, Yao et al. \cite{yao2024taubenchbenchmarktoolagentuserinteraction} constructed scenarios with API tools, policies about the tools, and LLM-simulated users to assess multi-turn interactions with language models. Jimenez et al.\cite{jimenez2024swebench} utilized Github issues and merged pull requests to construct a benchmark that evaluates models’ ability to resolve coding problems that exist in real world settings. These scenario-based approaches are promising for incorporating real-world contexts into model evaluations as the cost of constructing datasets for these evaluations is relatively low because they either scrape the web for data, adopt existing datasets, or create synthetic data using LLMs. However, due to the lack of user input in these datasets, this approach falls short in validating constructed scenarios as realistic downstream use cases or in evaluating real-world domains where there is no open-sourced practice-centered data.

Another line of research takes a more human-centered approach towards situating benchmarks in downstream usage by eliciting rich descriptions of benchmark design decisions or incorporating domain expert insights in creating benchmarks. An example of the former adopted evidence centered design (ECD) for designing assessments from educational tests and provides a framework for thinking through and recording design decisions during the process of constructing a benchmark \cite{liu2024ecbdevidencecenteredbenchmarkdesign}.
In the latter category, researchers have worked with practitioners in domains where we have seen a rapid uptake of generative AI technologies, such as law and medicine, to construct evaluation approaches that incorporate practical contexts. Guha et al. \cite{guha2023legalbenchcollaborativelybuiltbenchmark} formed a team of natural language processing researchers and legal professionals and crafted tasks to measure legal reasoning in generative models. Reddy et al. \cite{reddy2021evaluation} formulated evaluation guidelines for implementing AI in clinical settings based on existing frameworks and literature in medicine, and Hamna et al. \cite{hamna2025buildingbenchmarksgroundup} worked with community organizations to construct evaluations of language models in health chatbot settings.
These human-centered design-based approaches are effective for creating evaluations that are representative of real-world usage because they center the perspective of generative model users to highlight the usage context, and they encode human perspectives into a technical system via a design process. 

Our work builds upon these existing examples of human-centered design-based approaches to constructing evaluations that are situated in the real-world contexts. Specifically, we focus on soliciting practitioners' engagement in designing evaluations for applications of generative models in journalism. Previous work has already investigated how journalistic perspectives could be incorporated into designing a generative language model \cite{tseng2025ownership-8b2}, or evaluating generative models for journalism tasks \cite{li2024newsbench-770, patwardhan2025gdpvalevaluatingaimodel}, though none have take an explicit design-based approach to model evaluation in the domain. We instantiate a human-centered approach by iteratively engaging a diverse and substantial group of journalism practitioners throughout the process of designing and evaluating an illustrative benchmark. This benchmark design method enables us to capture real world usage and domain expertise, in turn allowing us to prototype a domain-specific benchmark rooted in the practices, values, and ethical reasoning of the domain \cite{diakopoulos2023leveraging-af4}.

\begin{table*}[h!]
\centering
\begin{tabular}{l l ll}
\hline
\textbf{Participant}& \textbf{Role} & \textbf{Organization Type}  &\textbf{Years of Journalism Experience}\\
\hline
P1  &  Director&  Non-profit and Public Media &More than 5 years\\
P2  &  Investigative Reporter&  Legacy Media &1-5 years\\
P3  &  Newsroom strategist&  Legacy Media &More than 5 years\\
P4  &  Professor&  Academic &More than 5 years\\
P5  &  Engineer&  Legacy Media &Less than 1 year\\
P6  &  Chief Product Officer&  Non-profit and Public Media &More than 5 years\\
P7  &  Lecturer&  Academic &More than 5 years\\
P8  &  Professor&  Academic &More than 5 years\\
P9  &  Director of Engineering&  Digital Media &1-5 years\\
P10 &  Media Relations Lead &  Software Company &More than 5 years\\
P11 &  Data Reporter&  Legacy Media &More than 5 years\\
P12 &  Newsroom strategist&  News Agency &More than 5 years\\
P13 &  Professor&  Academic &More than 5 years\\
P14 &  Consultant&  Start-Up &More than 5 years\\
P15 &  Data Editor&  Non-profit and Public Media &More than 5 years\\
P16 &  AI Strategist&  Local Media &Less than 1 year\\
P17 &  Editorial Director&  Legacy Media &More than 5 years\\
P18 &  Director of Technology&  Legacy Media &More than 5 years\\
P19 &  Director of Technology&  Legacy Media &Less than 1 year\\
P20 &  Director of an Academic Program&  Academic &More than 5 years\\
P21 &  Engineer&  Legacy Media &1-5 years\\
P22 &  Technology Editor&  Legacy Media &More than 5 years\\
P23 &  Graphics Reporter&  Legacy Media &More than 5 years\\
\hline
\end{tabular}
\caption{Participant Information}
    \label{tab:participant}
\Description{Table with 4 columns. Column 1 lists participant numbers; column 2 lists participant role; column 3 lists participant's organization type; and column 4 lists years of experience participant has had in journalism industry.}
\end{table*}

\section{Designing a Benchmark for Journalism with Practitioners}
To help address critiques surrounding the realism of generative model evaluations, we adopted a co-design workshop approach where we engaged journalism professionals in semi-structured group-based discussions about their experiences with using and evaluating generative AI in their work and considerations for creating a benchmark evaluation for journalism \cite{10.1162/DESI}. We trained a handful of researchers to facilitate and record notes observing journalism professionals in groups, and we analyzed these notes for challenges and tensions during the collaborative benchmark design process \cite{doi:10.1177/16094069241289278}. The co-design method was simultaneously beneficial for centering practitioners' perspectives in defining and creating evaluations of AI systems for research purposes and for building a community of practitioners who were invested in the impact of AI systems in the industry. We chose to use an observational approach, as opposed to audio recording, to enable our participants to partake freely without concerns regarding disclosure policies. In the following, we describe the workshop organization as well as findings and related design challenges.

\subsection{Workshop Organization}
To understand existing approaches and challenges to AI evaluation in newsroom settings, we helped structure and rapporteur a day-long professional workshop hosted by a large U.S. based research university. This workshop was a standalone event, and participants were not financially compensated nor required to pay to participate in the workshop. Some participants who were not local had their transportation and accommodation paid for by the host university.

\subsubsection{Participants}
Twenty-three journalism practitioners participated in the workshop (Details in Table \ref{tab:participant}). We recruited these participants via a purposive sampling method \cite{campbell2020purposive}, where members of the research team identified and invited individuals in the field of journalism who had experience working with or developing generative AI tools for news. We intentionally selected individuals to represent a range of organization types (from legacy media to academic institutions), roles and expertise (from reporting and editing, to product design and management), and years in the field.  In addition to the invited participants, six researchers received training beforehand and functioned as rapporteurs at the workshop, moderating discussions among the participants and taking notes on discussions. Consent was obtained from all participants for researchers to observe and analyze written artifacts and discussion notes from the workshop.

\subsubsection{Workshop Structure}
The workshop was organized into large group sessions and breakout sessions (see detailed session structure and participants in Appendix \ref{appendix:a}). In an introductory large group session (which we refer to as INT in the rest of the paper), participants were presented with definitions and an overview of existing benchmark research to level-set their knowledge. They were also prompted to discuss existing practices and challenges for generative AI evaluation in their respective organizations. Participants were then organized into several smaller breakout sessions consisting of 3 to 4 participants plus the rapporteur. Each individual participated in two 90-minute breakout sessions on different topics. After all of the breakout sessions, a discussion (referred to as SYN) was conducted with the whole group of participants to collectively synthesize challenges encountered during the exercises at the workshop and outline possible next steps.

To inform the content of each of the 90-minute breakout sessions, we distributed a survey prior to conducting the workshop to elicit importance ratings of benchmarking model performance for an array of journalism tasks and values on a scale from 1 (not important) thru 7 (important). Based on ratings collected from the survey and findings in previous research \cite{10.1145/3715336.3735717, 10.13140/RG.2.2.31540.05765, cools2024uses-635, diakopoulos2023leveraging-af4, Deuze2005if, Hanitzsch2007it}, we selected six news production use cases: Information/Data Extraction (UC1), Semantic Search (UC2), Summarization (UC3), Content Transformation (UC4), Background Research (UC5), and Fact Checking (UC6); and six values that are typically important to the practice of journalism: Accuracy (V1), Transparency (V2), Confidence/Uncertainty, (V3) Accountability (V4), Objectivity/Bias (V5), and Timeliness/Recency (V6) to ground discussions about creating evaluations within the news domain (see descriptions of each use case and value in Appendix \ref{appendix:a}). Participants were placed in groups corresponding to these selected use cases and values based on their responses to the survey, and complementary perspectives and backgrounds, and they worked through a structured set of discussion questions with respect to these use cases and values. 

The questions that participants worked through were somewhat different for the use cases and the values. For use case sessions, the questions included: \textit{How have you used generative AI for this use case? How would you assess the performance of specific tasks in this use case? And How would you assess the performance of specific tasks in this use case?}. For values sessions, the questions included: \textit{How is this value reflected in your work?, How do you imagine a generative AI system could violate this value? And How would you set up a test case to assess the alignment of a generative AI model or system with this value?}. Groups were provided with further discussion prompts that aligned with each guiding question (See Appendix \ref{appendix:b}). The groups were structured such that individual participants had a few minutes to contemplate and take notes for each question in turn before engaging in discussion with the rest of the group. 

\subsubsection{Data Collection and Analysis}
\begin{figure*}[ht]
\centering
\subfloat[\centering]{\includegraphics[width=.3\textwidth]{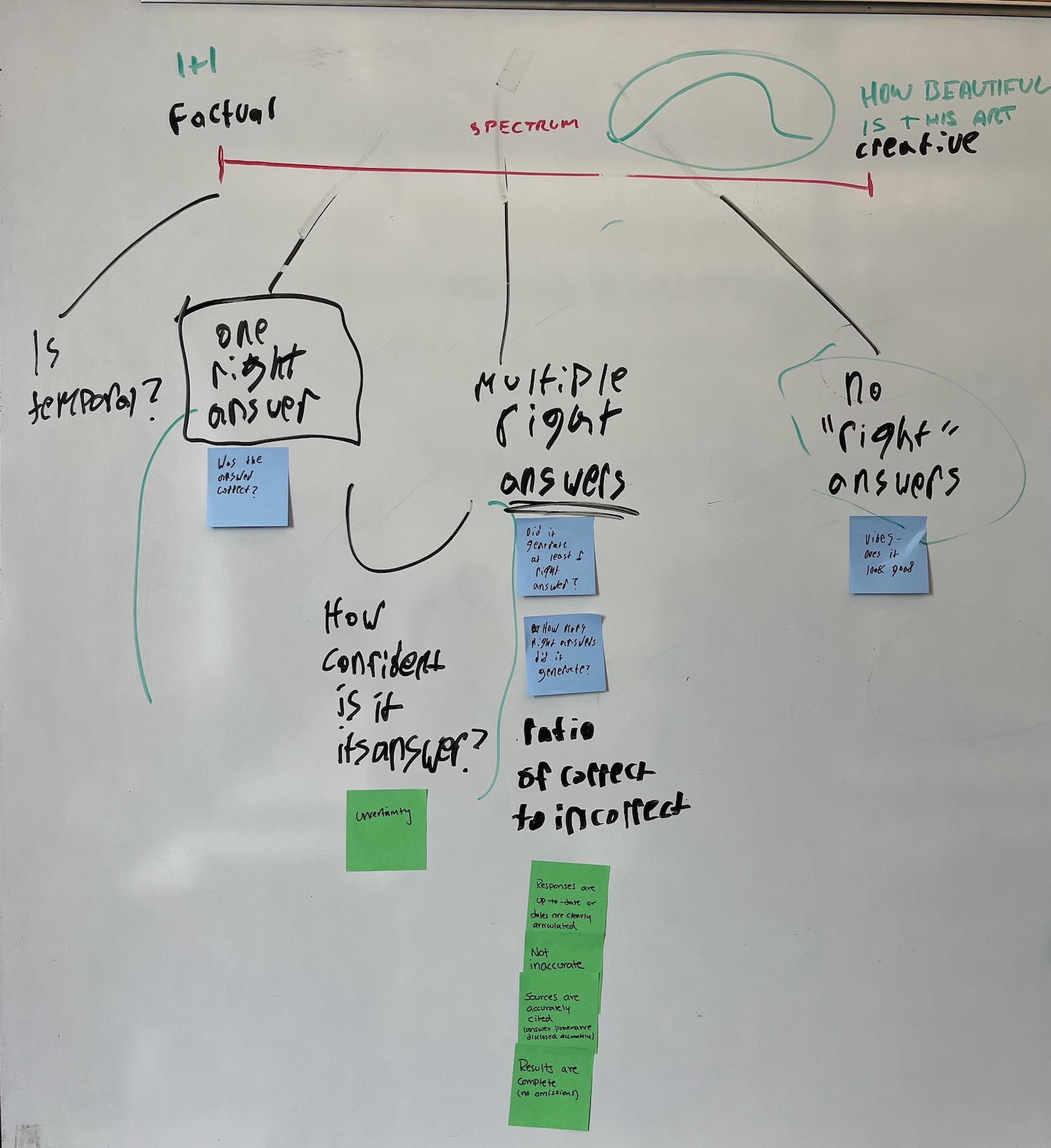}}
\qquad
\subfloat[\centering]{\includegraphics[width=.3\textwidth]{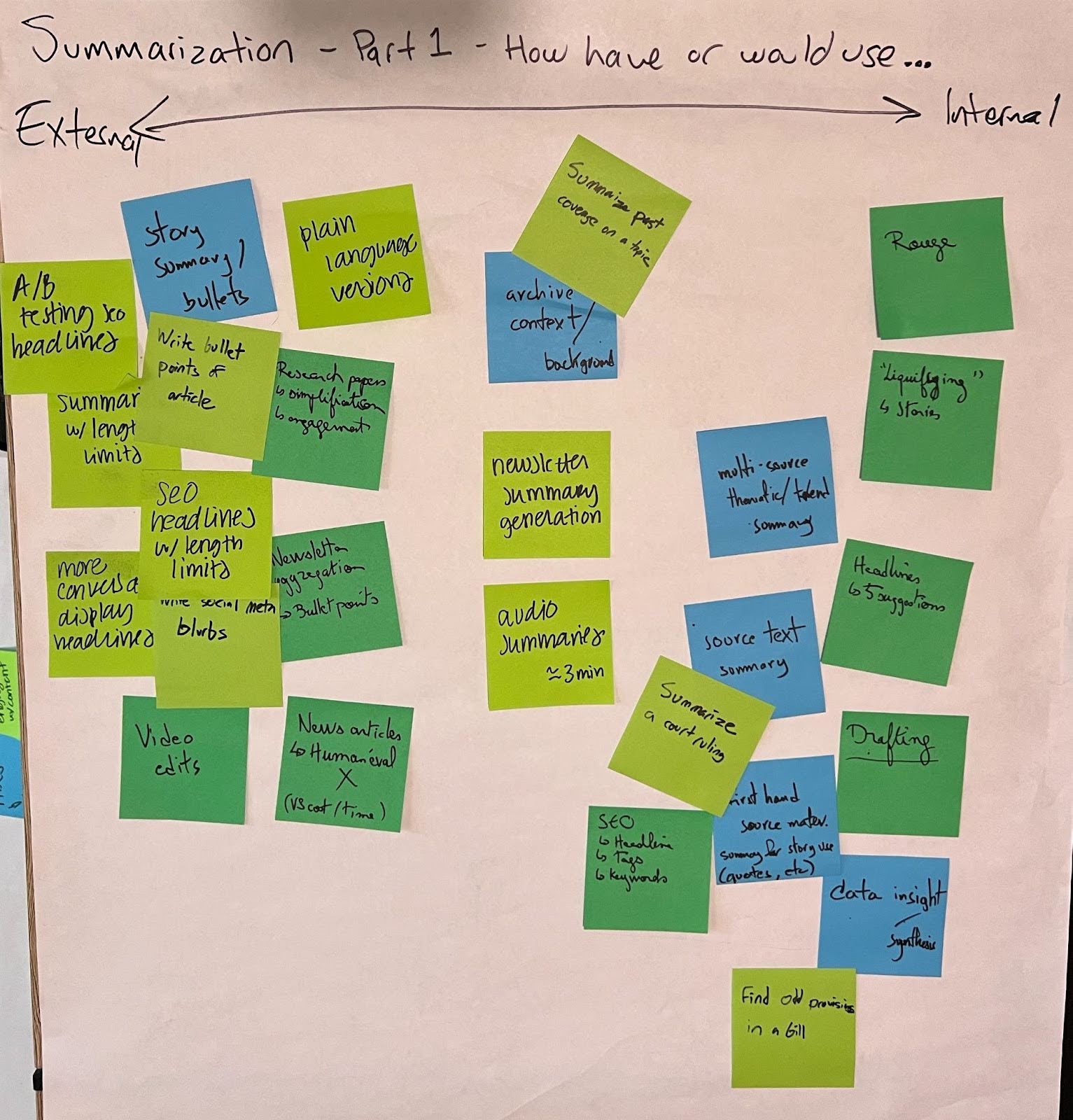}}
\qquad
\subfloat[\centering]{\includegraphics[width=.3\textwidth]{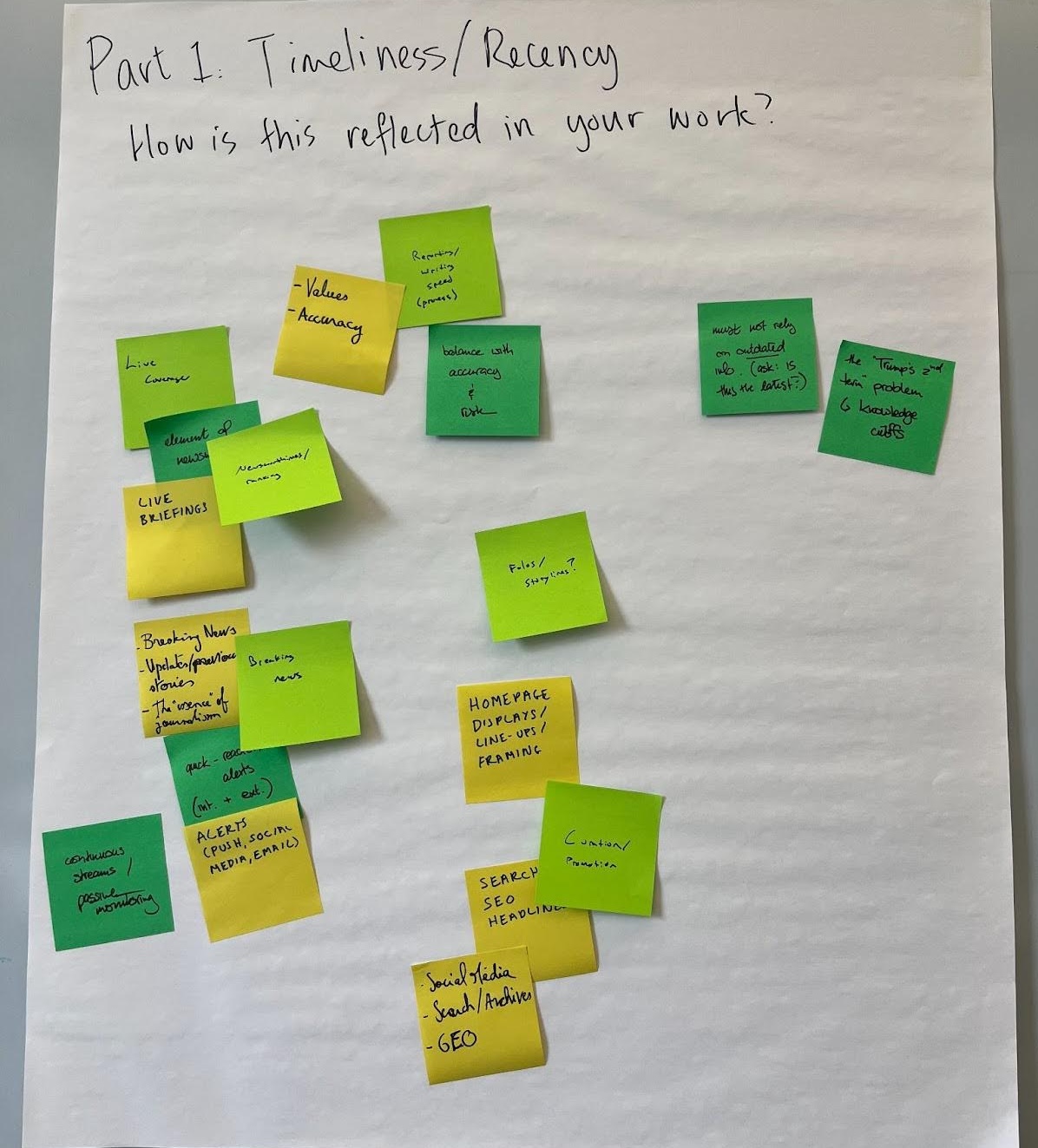}}
\caption{Participants wrote on Post-it notes their responses to discussion questions and collectively organized the notes on whiteboards. (a) shows notes for \textit{How to assess the performance on specific tasks in this use case?} in the Background Research session; (b) shows notes for \textit{How have you used or would use generative AI for this use case?
} in the Summarization session; and (c) shows notes for \textit{How is this value reflected in your work?} for the Timeliness/Recency session. }
\label{fig:postit}
      \Description{Three side-by-side photos of whiteboards with Post-it notes and scribble, labeled (a), (b), and (c) accordingly. Post-it notes are loosely organized in clusters in each image.}
\end{figure*}

Two types of data were collected during the workshop. For one, participants were encouraged to jot down answers to discussion questions on post-it notes and arrange them on white-boards during small group discussions (As shown in Figure \ref{fig:postit}). For another, we trained a group of six researchers, acting as rapporteurs, to moderate conversations among participants in group discussion and conduct direct participant observations of these conversations \cite{Ciesielska2018}. Specifically, we instructed the rapporteurs to note down perspectives, tensions, and consensus in participants' verbal discussions based on their subjective interpretation and to maintain participant anonymity in their notes. The breakout sessions thus resulted in 12 sets of digital notes taken by rapporteurs with accompanying photographs of participant-produced written materials. The first author also produced observational notes from the initial and final full group discussions. All six rapporteurs had prior experience in qualitative research and possessed in-depth knowledge about technologies in the journalism domain, which qualified them for the rapporteur role. 

The research team, with the help of rapporteurs, conducted a reflexive thematic analysis of the data we collected \cite{braun2006using}. Immediately after the workshop, the rapporteur team met to review all of the data collected during the workshop, noting themes that came up across several sessions and any challenges in moderating sessions, which also helped with familiarization of the data. The first author then labeled the collection of rapporteur notes with descriptive codes, which were clustered and used to refine themes discussed by the rapporteur team. The first author created memos regarding these themes and reported them back to the rapporteur team for feedback. Further feedback was integrated into memos and used to develop a final set of themes which we present next. 

\subsection{Workshop Findings}
\label{findings}
Our analysis of rapporteur notes and observations at the workshop surfaced four aspects where journalism practitioners experience resistance in using or creating evaluations of generative models that capture their professional knowledge. For each aspect, we describe realistic considerations that go into evaluating one's work, provide descriptions of goals practitioners would like model evaluations to achieve, and characterize challenges practitioners face with creating or using existing evaluations. We then lay out design challenges related to each aspect in creating evaluation systems.

\subsubsection{Context Specificity and Generalizability}
\label{context}
A general observation shared by all six rapporteurs moderating different discussions was the tension between participants caring deeply about in-situ evaluation of AI models for specific tasks within the scope of a reporting or development project and coming up with a broad framing and measurements for ``generic'' tasks in journalism. While some participants argued that a general journalism benchmark could act as a lever on model developers and a measuring stick for decisions about what models to use, many found it difficult to translate project-specific and ``vibe-based'' evaluations hinging on journalistic instincts to scalable and generalizable metrics that represent journalism at large. 

Participants provided several contrasting task contexts where different considerations were applied to evaluating AI models. For example, participants in the summarization session (UC3) noted that while engagement and text length were important metrics for evaluating summarization outputs served directly to an audience, they might not be applicable for summarization outputs used by journalists during reporting processes. And metrics like bias and source attribution would matter more for the latter. (A similar distinction between \textbf{internal and external usage} of model outputs was made in the fact checking breakout, UC6.) Participants also noted \textbf{news story types} (e.g. fact-based versus narrative-based) as a factor that required different evaluation metrics --- one required narrative tension while the other called for accurate representation of facts, not to mention there also existed stylistic differences between individual journalist's summary writing. Other factors that influenced the applicability and priority of different evaluation metrics included \textbf{input document's modality} and \textbf{size} (UC2, UC5), the \textbf{stage in editorial processes} (UC5), or \textbf{specificity of desired outputs} (e.g. searching for a specific date or document versus discovering as many relevant documents as possible) (UC4, UC5). Furthermore, task contexts also informed the input given to models, such as \textbf{negative prompts} and \textbf{instruction tuning}, or \textbf{iterations on prompts} (zero-shot versus few-shot) based on in-situ assessments of model outputs (UC1, UC2). Without these details describing the context in which an evaluation is conducted, practitioners were less likely to be convinced by reported scores representing model performances for any use case.

\textit{Design Challenge}: \textbf{Balancing Context Specificity and Generalizability}. The importance of task-specific contexts for evaluating AI systems among journalism practitioners suggests a need to capture and represent those contexts as a part of the evaluation design process. This could take the form of mapping out dimensions of variation of tasks or ways in which domain values are applied in context-specific ways. Representing contexts in the benchmark could enable a more comprehensive and systematic scope of a benchmark that facilitates generalization, or enables practitioners to calibrate benchmarks with their specific contexts of use.

\subsubsection{Values as Evaluation Metrics} 
\label{value}
Journalistic values have a multitude of meanings in the context of a newsroom. Most commonly, values help journalism practitioners judge what goes into a news story (i.e. what is ``newsworthy''\cite{Harcup2016bn}), but values such as accountability also define a standard for professional practices that go beyond writing a story \cite{diakopoulos2023leveraging-af4}. We intentionally left the definition of values vague in structuring the workshop in order to observe how practitioners contend with the complexity of journalistic value in regards to usages of generative models.

Because journalistic values were used to guide editorial processes in news, it was not surprising that participants naturally leaned into values when discussing criteria of success for generative model use cases in newsrooms. Specifically, participants referred often to \textbf{accuracy}, as measured through the factuality and recall of model outputs (UC1, UC2, UC3, UC5), and \textbf{uncertainty}, which practitioners saw as the degree of trust or skepticism about claims given available evidence and measured in language models through a model's self-report of output relevance (UC2, UC5). \textbf{Efficiency} as an evaluation metric also emerged in the sense of time saved (UC6) or human time needed in order to refine LLM outputs for publication (UC4). Some participants also referred to audience response (such as user satisfaction, feedback, and traffic to specific articles) as a metric for evaluating user-facing editorial outputs, placing an emphasis on \textbf{engagement}, a value related to building a connection with one's readership (UC4, UC6). Other values that were mentioned in discussions about evaluating generative AI systems on journalism use cases included \textbf{objectivity/bias} and \textbf{diversity} (UC2). For instance, in the semantic search session (UC2), \textbf{diversity} was defined by a participant to be a measurement for surfacing a mix of search results that are different from each other but still relevant to a search query.

While the kinds of values used by journalists to define the success and quality of task completions fell into the same categories, the importance of adhering to each value differed across different usage contexts and stakeholders. For instance, the definition of \textbf{accuracy} could differ between different scenarios: accuracy for newsroom internally-facing tools could be lower than externally facing tools (V1). 
In the context of semantic search (UC2), participants explained that accuracy would be defined in terms of relevance of search results but indicated that, depending on the task at hand, relevance assessments would sometimes be defined through calculating precision while in other times rely more on calculating recall. Additionally, the importance of adhering to accuracy could be significant from the perspective of news organizations and individual journalists because of legal and ethical reasons, but some participants indicated it might matter less for some audiences seeking local news from e.g. social media channels (V1). 

While some journalistic values worked in tandem with each other in practice, such as accountability and transparency (V2, V4), participants pointed out several tensions between adhering to many values as well. Timeliness could have a trade-off with values like accuracy and accountability where optimizing for efficiency of content creation might mean the lack of reviewing and validation from humans, leading to lower accuracy and accountability (V1, V4). Valuing business incentives could also be in tension with values such as transparency and uncertainty. Participants pointed out that disclosure of internal practices can be damaging to stakeholders and business models (V2), or when competing with other news producers who make bold news statements, overly confident disclosures might be incentivized (V3).

Lastly, participants also noted that, in some cases, using generative model-based tools outright undermined certain journalistic values. A participant pointed out the conflict between existing language models and the value of transparency by stating ``the process by which an LLM gets an answer is inherently not transparent [and] nobody understands exactly how these things work'' (V2). Participants also pointed out the tension between the value of accountability and the technology because the lack of reproducibility of language model outputs when using the same prompt makes it difficult to hold models accountable (V4). 

\textit{Design Challenge}: \textbf{Capturing Changing Values and Tensions}. Domain-specific values such as transparency, accuracy, objectivity, and others often function as metrics for evaluation as they reflect aspects of model capabilities that are important to practitioners. A challenge in designing evaluations with domain-value-driven metrics lies in capturing the fluid definitions and importance of values (and their tensions) across different stakeholders and usage contexts, a task which can benefit from work and methods on value-sensitive design \cite{friedman2017survey-119}.

\subsubsection{Dataset Construction}
\label{data}
All of the workshop participants acknowledged the need for high quality datasets that were representative of journalism tasks and values during the synthesis session (SYN), though they uncovered several tensions in identifying and creating such datasets for evaluation purposes. One tension faced by practitioners regarding dataset creation was related to privacy and transparency, with an additional layer of timeliness that was specific to journalism. On the one hand, when prompted with questions regarding data sources, one rapporteur noted that participants were more comfortable using internally-available data for evaluation as it reflected use cases that practitioners cared about (V5), and participants also voiced concerns surrounding technology companies adopting any published data for training purposes and gaming benchmarks (V6), which might result in data contamination (UC2) and further discourage open-sourcing evaluation datasets among journalism practitioners. On the other hand, participants worried about transparency and validity of evaluations based on internally-available data advocated for the use of publicly-available data, though in addition to concerns about data contamination, participants still raised concerns about the ethics with using data that was not one's own and timeliness of public data since news reporting routines could be dynamic (V1). 

Regardless of access to existing data, high quality data was critical for journalism-specific evaluations of generative AI, and constructing such data was an essential next step for benchmarking generative AI with regards to journalistic work. In several discussions, participants mentioned that human feedback and validation was required for identifying and constructing ground truth datasets for evaluation (UC1, UC4, V1, SYN), such as when a participant raised the question: ``How to create benchmark datasets that are less intensive?'' during the discussion about where to source data for a journalism benchmark in the synthesis session. While the need for datasets to incorporate domain-expert knowledge was a well-known challenge in benchmark research and have been addressed in benchmarks with expert annotators like GPQA \cite{rein2023gpqagraduatelevelgoogleproofqa}, the challenge for constructing journalism-specific datasets lay in various possible ground-truths and stakeholders. When prompted to come up with evaluation datasets, participants stated that there could be stylistic differences among newsrooms and reporters (UC3). Other participants said ground-truth could also be dependent on use case (UC4), or values like accuracy could not be objectively defined (V1), reflecting challenges in the previous subsections. Participants also raised the question ``who is considered an expert?'' in several cases (UC3, UC4, V1), mentioning the importance of audience, editorial staff, or the legal team in determining a gold standard for different use cases and values across various small group discussions. Moreover, one of the most critical resources for dataset creation was time and human labor, which participants voiced to be lacking at the organizational level, creating an additional barrier to constructing satisfactory evaluation datasets (INT, SYN).

\textit{Design Challenge}: \textbf{Protecting Data Ownership while Facilitating Accessibility}. As stated above, the lack of professionally validated datasets is a well-known challenge within generative model evaluation. Our workshop surfaces a specific tension in journalism surrounding data ownership, privacy, and accessibility, which could apply to other professional domains as well. Addressing this challenge might require the research community to shift attention to understanding the power dynamic between data producers (such as newsrooms in this case) and data users, providing protection and empowering data producers through policy, and incentivizing the open-sourcing of data.

\subsubsection{Variations in Human and Organizational Factors}
\label{organization}
Lastly, discussions at the workshop also pointed to several organizational and human factors as sources of many aforementioned challenges and tensions in creating realistic generative model evaluations that reflect journalistic work. One individual-level factor that required attention from benchmark designers was rooted in technical fluency of end users. In addition to task-contextual factors that might dictate the way users interact with and create prompts for a generative system, personal technical fluency could impact how users interact with a generative system in terms of taking multiple turns with a model, entering and tuning with prompts, or interpreting outputs and utilizing outputs from a model. Specifically, participants mentioned the need to account for different prompts from novice users and expert users when benchmarking (UC2). Another participant said that in their work process, they often utilize ``lazy quantitative eval[uation]'' on a small subset of examples in order to adjust subsequent prompts (INT). In fact, most in-situ evaluations of generative AI systems mentioned by participants in the full-group discussion had to rely on individual, story-specific evaluations rather than more generic and already produced model benchmarks. In other words, they wanted to know how well the model worked for the exact task they had.

Organizationally, participants mentioned the importance of commercial aspects of news and audience engagement, which was also reflected in considerations for creating value-based evaluation metrics. It was important for participants that the use of generative models could enable them to reach a broad audience (UC3) and compete with other organizations or other news formats (UC4),  Furthermore, aspects like abiding by legal policies and organizational policies could also differ organizationally to the extent that a participant posed that ``[journalism is] not a uniform industry for bias, or accuracy,'' (UC3) highlighting possible subjectivity across organizations in defining values and hence metrics for evaluations.

These differences among individual journalists and journalism organizations also surfaced in discussing the utility of benchmarks and evaluations in journalism. Specifically, participants were varied in support for collecting data and creating an all-encompassing benchmark that was representative of journalism versus synthesizing a framework that could be adapted for conducting evaluation in individual scenarios. Those in support of a standardized benchmark believed benchmarks could help determine whether to use a new model or not, and that to operationalize such a benchmark, the industry should collectively invest in canonical data for training purposes and releasing them (SYN). Conversely, many participants felt that efforts should be put into ``making a framework for developing benchmarks, [such as a] cookbook for evaluating in a newsroom instead of a set of data for people'' (SYN). This idea was also present in breakout sessions (UC4, UC5): for example, in the content transformation session, participants wondered if there might be a higher-order framework or dedicated set of resources that could help shoulder conducting evaluation across many different formats and modalities. Therefore, it is necessary to reflect critically on the value of evaluation for an industry with diverse preferences and standards like journalism.

\textit{Design Challenge}: \textbf{Supporting Modularity and Adaptability}. Due to individual and organizational variances for evaluation including everything from business objectives, to prompting expertise, and the definition of metrics, a challenge to designing a journalism-specific benchmark lies in accounting for varying needs and goals of evaluation across individuals and organizations. This challenge calls for a design that is informative for users regarding utilities of evaluations and is flexible, adaptable, and customizable to different evaluation needs. Modularity and adaptability would allow users to define some standardized tasks and then select and refine datasets as needed according to individual or organizational goals.

\section{Prototyping a Journalism Benchmark}

Motivated by the discussion surrounding creating a framework that could be adapted into individual evaluation context as opposed to collecting canonical data for a generalized journalism benchmark (\ref{organization}), we introduce a prototype of a journalism evaluation cookbook. We designed this evaluation cookbook to help ground challenges and tensions mentioned during the workshop in the process of design and used to further reflect on the design challenges surrounding creation of evaluations that are responsive to real-world use cases. In the next subsection we describe the cookbook design, in section \ref{design_considerations}, we connect the design back to the design challenges as described in the previous section (\ref{findings}), and then in section \ref{design_reflection} we further reflect on the design process. 

\subsection{Cookbook Design}
\begin{figure*}[h]
  \centering
  \includegraphics[width=0.7\linewidth]{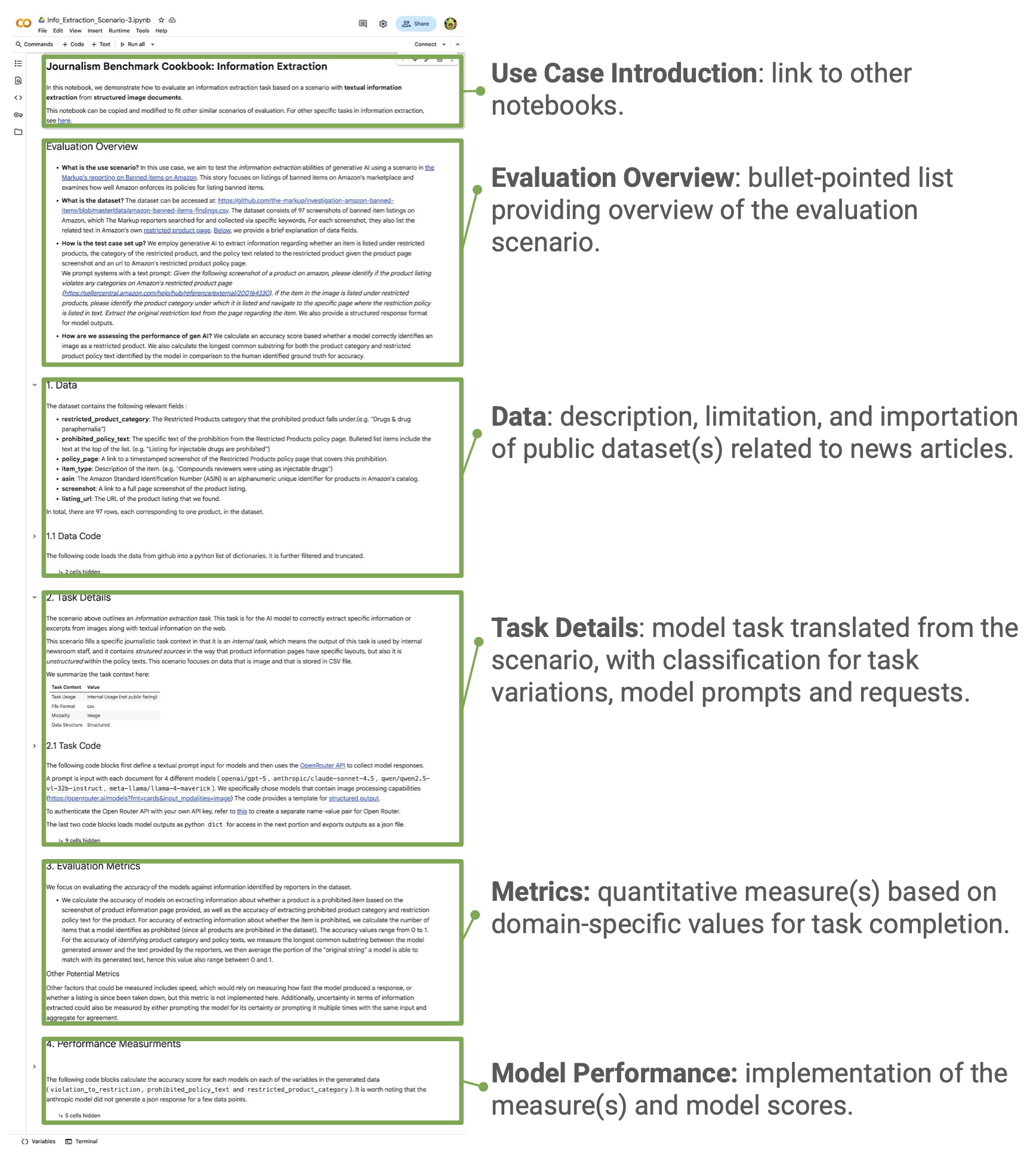}
  \caption{Google Colaboratory Notebooks are modular and allow us to interweave markdown text and code explaining rationales behind decisions in the evaluation process.}
  \label{fig:notebook}
  \Description{A screenshot of a Google Colaboratory Python Notebook. The file name is "info_extraction_scenario-1.ipynb," and it shows paragraphs in Markdown, starting with a title "Journalism Benchmark Cookbook: Information Extraction" and some bullet pointed text providing an "evaluation overview."}
\end{figure*}

The prototype consists of three Python Notebooks implemented in Google Colaboratory (Fig. \ref{fig:notebook} is an example of one of these notebooks and demonstrates the overall structure of an evaluation notebook. For details in the notebooks, refer to Github repository.\footnote{https://github.com/shallotly/news-eval-cookbook}) Each of the notebooks corresponds to an information extraction task based on a published news article and the associated dataset reporters used to produce the story. The news articles and associated datasets included are (1) \textit{How federal agencies responded to our requests about AI use in FOIA}\footnote{\url{https://www.muckrock.com/news/archives/2025/may/07/how-federal-agencies-responded-to-our-requests-about-ai-use-in-foia/}}, (2) \textit{Amazon’s Enforcement Failures Leave Open a Back Door to Banned Goods—Some Sold and Shipped by Amazon Itself} \footnote{\url{https://themarkup.org/banned-bounty/2020/06/18/amazons-enforcement-failures-leave-open-a-back-door}}, and (3) \textit{Apple says its app store is `a safe trusted place.' We found 1500 reports of unwanted sexual behavior on six apps, some targeting minors.} \footnote{\url{https://www.washingtonpost.com/technology/2019/11/22/apple-says-its-app-store-is-safe-trusted-place-we-found-reports-unwanted-sexual-behavior-six-apps-some-targeting-minors/}} Each notebook is designed to consist of 5 components: 
\begin{itemize}
    \item \textit{Evaluation Overview}. This component is presented in the beginning on the notebook in a bullet-pointed list, which provides a brief overview of the use case scenario regarding how the test dataset is produced, how the test cases for generative models are set up, and how metrics are applied. It is important to provide this overview in the beginning of the document because it not only provides necessary details in an accessible language that help a user decide whether the evaluation is relevant, but it also demonstrates how a use case scenario gets translated to data and metrics for testing a specific journalism task. 
    \item \textit{Data}. This component provides a more detailed overview of the dataset used in the evaluation, detailing a data dictionary, basic data summaries, limitations and missing data. It also contains a subsection with code for downloading and preparing the associated dataset. The main purpose of this component is to provide an example public dataset that represents the journalism task being evaluated. 
    \item \textit{Task Details}. This component introduces the computational task for generative AI models based on the use case scenario in the notebook. In a table, it outlines different contextual variabilities this task addresses (e.g. input file format, audience, modality, etc.) and provides code for constructing prompts, running the prompt on the given dataset via API calls, and recording model outputs. The prompts are designed based on methodology descriptions provided in GitHub repositories by story editors and defined in a stand-alone cell for easy access and modification by the user. Additionally, it also provides a rough estimate of the financial cost to run the evaluation. The code block contains separate cells for the prompt and a list of model names, allowing for easy modification of the prompt and selection of models to test.
    \item \textit{Metrics}. In this component, we explain the specific values we design metrics around. We name values such as accuracy and explain how we rate model outputs in comparison to the ground truth to calculate score(s) that represent such values. Regardless of whether we implement the calculation or not, we describe potential values to measure in the output of the model. 
    \item \textit{Model Performance}. This component contains code for implementing the metrics in the last section and neatly presenting the performances of different models on each of the metric, enabling easy comparison across different models. 
\end{itemize}

These notebooks are intended to be used by journalism practitioners who are interested in evaluating generative models in their work. We imagine these cookbook examples can be used in many different ways depending on the level of expertise and investment a practitioner has surrounding generative model evaluation. On one end of spectrum, these notebooks run successfully as is, which could be an informative overview of defining metrics and implementing an evaluation. On the other end of the spectrum, we also imagine practitioners with their own internal data could borrow the bare bone structure of the notebook and replace details, such as data, prompts, and metrics of the evaluation. Additionally, the structure of the notebook might serve as a template for reporting evaluation ``recipes''--evaluations that practitioners have done on their own. Our hope ultimately is for the cookbook to gain purchase across the domain, which could result in a wide range of different notebook-based evaluation implementations with datasets that might eventually be aggregated into a benchmark representative of journalism at large. Rather than assuming that a benchmark designer can fully capture all the necessary context to make an entire benchmark representative and valid, this approach enables domain practitioners to capture and reflect that context through a distributed and decentralized process.

\subsection{Design Considerations}
\label{design_considerations}
In this section, we explain how we operationalized challenges surfaced in the workshop through design decisions in the prototype and detail areas where we experienced friction in the design process.

\subsubsection{Addressing Context Specificity}
To address the challenge of specificity, we designed the cookbook prototype based on published news stories reflecting realistic news production scenarios related to information extraction tasks. We browsed public newsroom code repositories in order to select published data-driven news stories with data annotated and validated by reporters and descriptions of the reporting process to understand how datasets were used in a news setting. We specifically selected datasets that represented different story production contexts, reflecting varying input file formats (.pdf, text file), input modalities (e.g. text vs. image), and levels of structure in data, and we highlighted these contextual differences in the Task Detail component of each notebook to contrast how evaluations of the same use case (information extraction) can vary along various dimensions. However, given three notebook examples within one use case, we do not claim to have comprehensively listed all dimensions where the context of evaluation could differ in journalism. For example, one of the important dimensions of task context mentioned by participants was the intended audience of the task (i.e. whether the output is public facing or for internal use), but given the nature of information extraction in data journalism, journalists rarely presented information extracted from data directly to their audience, so we did not explore that variability here. Therefore, these dimensions in the prototype constitute a first step towards characterizing evaluation contexts for journalism tasks and aim to encourage practitioners to think about how their own evaluation might differ from these.

\subsubsection{Encoding Values}
Our approach to addressing the challenge of encoding journalism specific values in evaluation metrics relied on access to data that encapsulated journalistic judgments. Specifically, we tested for ``accuracy'' based on human-coded ground truth in all three datasets, and for one of the datasets we tested ``uncertainty'' by comparing models' response to ground truth entries where the human reported ``unsure.'' In the notebooks, we detailed how we compare model outputs to human-validated ground truth in the Metrics component in order to demonstrate how we translated journalistic value-based judgments to measurable metrics. This component aimed to help end users practice translating their professional values from instincts to concrete standards with explanation. In addition to providing sample metrics based on real data, we also noted, in these notebooks, other potential values to measure if we could obtain further human feedback, allowing users to conduct internal testing on metrics they care about.

\subsubsection{Data Access}
The challenge of accessing data for evaluations was an especially prominent one in the development of the prototype. In order to address the aforementioned challenges, we looked for journalism datasets that were varied in usage contexts and included professional judgments to serve as ground truth. The existence of professionally annotated and public-available news data in data journalism, despite not for the purpose of evaluating generative models, allowed us to address the challenge of data accessibility for the use case of information extraction \cite{Dodds08082024}. In the notebooks, we provided detailed descriptions of data properties in the Data component and constructed model prompts based on methodological descriptions that came with the datasets in the Task Details component. We also noted limitations in the data we obtained, including possible human errors in the ground truth data to enable users of the notebook to make their own interpretations of the evaluation. In our search for data we found only a small fraction of publicly available reporter-validated datasets for this use case despite the culture in data journalism to open source story data. The majority of publicly available data were not published with reporting methodologies or reporter annotation, making encoding values into model evaluation metrics impossible. Given challenges surrounding data access, the prototype also provided examples of how professional judgment was applied to evaluate the use case in the public datasets, which could serve as a road map and potential motivator for end users to apply their own (or their organization's) professional judgment towards contextualized versions of an evaluation.
 
\subsubsection{Adaptability}
 We attempted to address the challenges of adaptability through the format of the prototype design. By introducing the concept of the evaluation cookbook, which consisted of a series of computational notebooks, we aimed to provide an overall structure with use case scenarios, data, and metrics that could be copied and edited by practitioners to suit their own context. The notebook format interwove textual explanations, visualizations, executable codes, and benchmark results, allowing for varying levels of technical expertise to gain useful information or adapt it for their own evaluation \cite{OHara2015ComputationalNF}. We specifically isolated code blocks that users of all levels might modify, such as model prompts and lists of models tested, to allow engagement without breaking other parts of the code. Additionally, the notebook format was modular and allowed for easy modification and adaptation of parts of evaluation, as intended for a cookbook, where users should be able to modify the orchestration of evaluation freely.

\subsection{Reflection on Cookbook Design Process}
\label{design_reflection}

Overall, the process of developing the cookbook prototype for evaluating generative models in the journalism context reinforces several of the challenges surfaced by workshop participants. One of the most immediate struggles hinges upon constructing and accessing datasets. In addition to searching through open source journalism code bases online to find data for the prototype, we also reached out to participants of the workshop for the potential of sharing internal data, such as for the use case of summarization. While we received positive responses for sharing data for research purposes, no organization agreed to open source these internal datasets for a public benchmark. Moreover, our approach to obtaining open source datasets is only viable post (news) production, since reporting teams do not often publish their methodologies and edited datasets until after a news story is published. This reflects the difficulty of constructing realistic evaluations for a domain such as journalism as a designer without direct access to organization practices and data. Workshop participants expressed the need for in-situ evaluation and validation of model capabilities while news production is ongoing, which further highlights the need for designing frameworks that enable practitioners to define and conduct evaluations on the fly. Other challenges we faced as designers of a domain-specific benchmark included the difficulty in creating realistic prompts that reflect how a practitioner might use generative models in their work. We were able to make inferences based on the published methodology articles for the datasets we collected, but for datasets that do no have methodological descriptions or are not created for the purpose of testing generative models, it would be more difficult to make these inferences. Lastly, while the cookbook prototype accounted for some values that were encoded in datasets through reporter-coded data, we are not able to account for how different values might interact with each other in practice, and how these interactions might be meaningfully conveyed or traded-off in a tailored evaluation. 

\section{Discussion}
Through our observation and analysis of a professional workshop oriented towards designing practitioner-centered benchmark for journalism, this work responds to the question of \textit{What are
the needs and challenges surrounding the design of a domain-oriented generative model benchmark,} by presenting a set of design challenges in creating generative model evaluations that align with practitioners' downstream usage. These challenges include capturing variabilities across different task contexts to appropriately scope a benchmark (\ref{context}), articulating journalism specific values through evaluation metrics that can differ and contradict based on task contexts (\ref{value}), fostering a data ecosystem that acknowledges ownership and privacy of data providers (\ref{data}), and offering customizable evaluation structures that adapt to individual and organizational differences (\ref{organization}). We further elaborate on these issues through the design and prototyping of a benchmark cookbook aiming to address each challenge. 

Through these activities we are able to identifying a gap between the evaluation affordances of existing evaluation systems (such as generalized benchmarks) and the evaluation needs of journalism practitioners in their work. Based on this gap, we propose an orientation of focus for future evaluation designers in order to create evaluations that are grounded in real-world practices (\ref{benchmark_alignment}). Additionally, we reflect on our practitioner-centered design-based approach for collecting necessary components and constructing generative model evaluations and discuss potential applications of design in creating ecological valid evaluations (\ref{design_for_ecological_validity}) in order to address our question about \textit{How might practitioner-centered design processes address existing criticisms about validity in benchmark design?} Lastly, in section \ref{journ_consideration}, we detail additional journalism-specific factors to consider when creating technologies intended for use by journalism practitioners.

\subsection{Towards Evaluations for Practitioners} 
\label{benchmark_alignment}
Our work was motivated by prior research investigating the utility of existing benchmarks beyond machine learning development for end users. Hardy et al.\cite{10.1145/3708359.3712152} found that while benchmark scores might be used in academia to track research progress, their effectiveness in signaling model superiority and measuring downstream applications might be lacking due to their failure to reflect real-world uses. Schwartz et al. \cite{schwartz2025realitychecknewevaluation} also argued that existing evaluation approaches might measure what they call ``first-order effects'' such as the factual accuracy of immediate system outputs, but they might fail to account for indirect and ``second-order effects'' of generative systems (i.e. ``outcomes and consequences that may result from their use in the real world'') in different sectors of society such as education, healthcare, or employment. Our findings align with this prior research, showing that journalism practitioners infrequently rely on existing benchmarks for making decisions regarding which model to use for which task. Instead, they opt to measure the quality of AI models on specific tasks at hand and rely on ``vibe-based'' evaluations of immediate model outputs in order to iterate on model prompts until they obtain satisfactory results (\ref{context}). 

Yet our workshop findings also indicate that there is interest in the community of practitioners for informing evaluations of LLMs in a more rigorous way. In the workshop, we invited practitioners to think through example use case scenarios of AI and criteria they had used to determine the quality of outputs before asking them to imagine computational tasks and metrics that capture those criteria. In doing so, we observed that individual journalists might not be aligned with each other in terms of goals and preferences for evaluation in their day-to-day work---some hoped for an industry standardized benchmark that represents the perspective of journalism at large, while others wished for guidelines for thinking through measuring the performance of AI in specific scenarios (\ref{organization}). However, regardless of preferences between these possibilities, practitioners all conveyed a sense of urgency for some type of solution. 

While we find that access to data and annotated data by domain experts is a real gap, it is also more than that: it is about eliciting use case contexts, value tensions for metrics, and organizational or individual variations. Many existing evaluation approaches aim to capture domain expertise through data annotations and human feedback by domain experts \cite{guha2023legalbenchcollaborativelybuiltbenchmark} to address the research gap, but our research points to some alternative approaches. On the one hand, designers of domain centered benchmarks can consider a range of human-centered design methods for collecting rich contextual information so it can be reflected in benchmarks (\ref{design_for_ecological_validity}). On the other hand, stemming from a collaborative perspective between practitioners and evaluation developers, we envision an evaluation ecosystem that is created to inform the literacy and ability of end users in developing their own evaluations. In this case, resources and infrastructures are designed to help end users better understand the process of evaluation, and in turn enable them to adopt and conduct customized evaluations in their own contexts, with the potential that individual evaluation cases could be aggregated into a larger, more representative set of evaluations for the domain at large. Our design of the journalism benchmark cookbook prototype is partially informed by this idea of evaluation literacy as well, by providing detailed explanations to aid understanding of each step of evaluation and an overall structure for which journalism practitioners can follow evaluations of their own. 

Several lines of existing research could already benefit the design of evaluations that promote literacy among practitioners. For instance, Dhar et al.\cite{dhar2025evalcardsframeworkstandardizedevaluation} proposed a framework for standardized documentation and reporting of evaluations, providing standardization and transparency surrounding input format, evaluation prompts, metrics, and safety. A similar approach could be adopted for practitioner-centered evaluation both to allow easy interpretation of individual evaluation results and to standardize how practitioners might record their own test cases. Further research down this line might also investigate the sociotechnical requirements for implementing standardized documentation across an industry (such as journalism) where technical literacy varies. Another line of prior work that might also inform evaluation literacy lies in the concept of ``design in use.'' This concept focuses not only on the design of an initial system to be tailorable for downstream uses, but also engages with how end users modify the system to fit their own work situations as sites for design innovations \cite{HENDERSON1995793}. Constructing evaluation systems for practitioners through the lens of design in use entails designing not only an initial evaluation system that functions to capture contextual needs of practitioners, but also researching through how practitioners tailor and rebuild the design to fit their own needs as they might gain knowledge about how to conduct evaluation through the process. This research direction might also extend to examine how individual practitioner's expertise or familiarity with language models and benchmarks inform their perspectives and preferences for evaluating language models in professional contexts. Future research on a design in use process might further shine a light on the needs of practice-oriented expertise-specific evaluation of generative AI systems.

\subsection{Designing for Validity in Benchmarks}
\label{design_for_ecological_validity}
Our work responds to a body of work on generative model governance and participation in generative models. Tseng et al. \cite{tseng2025ownership-8b2}, which instantiated the work of Suresh et al. \cite{10.1145/3630106.3658992}, conducted a co-design study with journalists to envision how a journalist-led large language model would work and argued that a design-based approach to large language models could enable holistic and contextual evaluations, foster community-building, and offer educational value for end users. Our findings align with this argument in that we were able to foster a community of practitioners invested in evaluating generative models within their domain through collaborative design inquires, and this approach also allowed us to uncover disputes and variabilities within this community of practice. Hamna et al. \cite{hamna2025buildingbenchmarksgroundup} provided an example of incorporating community insights through design into the construction of a benchmark, though their participants consisted of civil-society organizations who provided inputs regarding the community members they served, which differed slightly from our approach where we worked directly with intended end users. However, as mentioned by our participants (\ref{value}), values behind evaluation metrics could involve various stakeholders including their audience, a community journalists serve, inviting the incorporation of stakeholders beyond practitioners in the design of benchmarks.

In light of the line of research on design-based methods for constructing and evaluating generative models, we consider our findings with respect to addressing validity in benchmarks through design. Firstly, through engaging journalism practitioners in discussions about designing evaluation with specific structure for use case and value, we were able to obtain a preliminary set of dimensions (internal versus external audience, input and output format, and etc.) to define contexts of downstream evaluation (\ref{context}). This understanding of evaluation context could help scope the specific set of constructs in a benchmark, which might serve as a prerequisite for frameworks like DICE \cite{shrivastava2025diceframeworkdimensionalcontextual}, and ultimately ground measurements in a benchmark within practical use cases described by contextual dimensions, resulting in ecological validity for a domain like journalism. Secondly, our approach uncovered several tensions in journalistic values in different evaluation contexts (\ref{value}) and suggested a gap in translating the fluid definitions of domain values into a set of metrics. This need to operationalize values from individual and contextual perceptions as technical measures invites a design-based approach \cite{10.1145/3715336.3735717}. While prior work in the benchmark community, such as HELM \cite{liang2023holisticevaluationlanguagemodels}, incorporated multiple measures informed by different values for generative models and provided the affordance for users to weight one metric over another, the set of values provided were fixed across all contexts. For creating benchmark with metrics grounded in ecological validity, designers might lean into value-sensitive design methods \cite{sadek2023designing} to understand how values are related to each other in different contexts, how values might change as specific technology develops over time, and how any particular model measurement encodes a value or a set of values. Lastly, design-based approaches might also be effective in addressing the issues of data ownership and accessibility as discussed in our workshop findings, in turn allowing the creation of benchmarks with datasets centered around downstream use. Specifically, benchmark designers might explore issues of data governance and privacy through design \cite{10.1145/1629175.1629210}, with the aim to provide infrastructure for data donation, redaction, and access management to counter issues of privacy, ownership, and benchmark contamination (\ref{data}). Future work might also investigate how design-based approaches could effectively be used to create ecologically-valid benchmarks in similar knowledge domains, such as academia, law, or medicine.

\subsection{Broader Field-level Considerations}
\label{journ_consideration}
Our findings highlight challenges in designing generative model evaluation systems that capture the real-world contexts in journalism, and we have shown how these challenges inform further approaches for both constructing evaluations that are useful for end users and designing benchmarks with ecological and construct validity. Here we offer a couple of additional considerations which capture issues that emerged as they intersect with the broader field and industry of journalism, namely the uneven application of practitioner attention in the design process and the competitive pressures and dynamics specific to the field. 

First, as our participants established during the workshop, editorial attention is an important resource in newsrooms, and harnessing attention and input from news professionals collectively is costly and difficult. Newsrooms are busy and time pressures make getting journalists' time and attention a challenge. Additionally, the topic of benchmarking is highly technical, and news professionals come from varied technical backgrounds and news practices. That makes it challenging to have technical group conversations. More importantly, resources among newsrooms are not evenly distributed. Smaller local newsrooms may not have the bandwidth and human resource to participate in shaping evaluation of generative models, although their work may benefit significantly from AI assistance \cite{russell2025aiuseamericannewspapers}. The condition of lack of editorial attention across the board and unevenly distributed human and technological resources might imply that the design of an evaluation system for journalism practitioners should account for the uneven landscape and needs in the industry by including appropriate beneficiaries and stakeholders in conversations or explicitly focusing on specific sub-groups of practitioners, such as local newsrooms or individual content creators.

Another important consideration when designing for journalism practitioners in the U.S. specifically is the larger media market dynamics and business structure of journalism, which was also highlighted in Tseng et al. \cite{tseng2025ownership-8b2}. Competition for revenue streams looms over journalism organizations, large and small, in the face of new forms of news consumption, such as social media and chat-based AI, exerting pressure on news practitioners to emphasize the value in audience engagement and retention over other journalism values \cite{Young20102024}. Deteriorating audience trust of news \cite{doi:10.1177/14648849241299775} requires more transparency and disclosure from news organizations than ever before. Missteps in news practices and business strategies could lead to grave failures. Together, these forces creates heightened business risks for news organizations to share and collaborate on resources such as internal data and the limited editorial attention when there is no immediate incentives to do so. Research in the future might consider ways in which to align agendas and build trust with organizations in order to form mutually-beneficial and long-term partnerships when considering the design of generative AI systems rooted in practitioners' needs.

\section{Conclusion}
In this work we present findings which help advance the design of generative AI evaluations that are responsive to the needs of practitioners in the journalism context. Through observations of a professional workshop of journalism practitioners we analyzed the needs, tensions, and challenges of designing evaluations for practitioners in the field of journalism. We then designed a prototype journalism evaluation cookbook which was responsive to these challenges but also helped us further reflect on the challenges as we developed it. Our discussion elaborates on the approach towards enabling the evaluation of generative AI for news practitioners, addressing validity issues, and considering broader field-level constraints. By studying these issues in the field of journalism we contribute specific insights around context specificity, domain values and tensions, and data access considerations that can inform the design of benchmarks for that field. Ultimately, our work helps show the value of taking a user- and domain-centered design approach towards developing generative AI benchmarks.

\begin{acks}
We would like to acknowledge staff members of the Knight Lab at Northwestern University for generously offering space and help for this research.
\end{acks}

\bibliographystyle{ACM-Reference-Format}
\bibliography{sample-base}

@String{Computing = "Computing" }

@String{Chelsea = "Chelsea" }

@String{Springer = "Springer-Verlag" }

@inproceedings{10.1145/3715336.3735717,
author = {Nishal, Sachita and Diakopoulos, Nicholas},
title = {Values as Problems, Principles, and Tensions in Sociotechnical System Design for Journalism},
year = {2025},
isbn = {9798400714856},
publisher = {Association for Computing Machinery},
address = {New York, NY, USA},
url = {https://doi.org/10.1145/3715336.3735717},
doi = {10.1145/3715336.3735717},
booktitle = {Proceedings of the 2025 ACM Designing Interactive Systems Conference},
pages = {2975–2991},
numpages = {17},
location = {
},
series = {DIS '25}
}

@techreport{10.13140/RG.2.2.31540.05765,
    author = {Diakopoulos, Nicholas and Cools, Hannes and Li, Charlotte and Helberger, Natali and Kung, Ernest and Rinehart, Aimee and Gibbs, Lisa},
    title = {Generative AI in Journalism: The Evolution of Newswork and Ethics in a Generative Information Ecosystem},
    institution = {The Associated Press},
    year = {2024},
    month = {04},
    doi = {10.13140/RG.2.2.31540.05765}
}

@article{Dodds08082024,
    author = {Tomás Dodds and Valeria Reséndez and Gerret von Nordheim and Theo Araujo and Judith Moeller},
    title = {Collaborative Coding Cultures: How Journalists Use GitHub as a Trading Zone},
    journal = {Digital Journalism},
    volume = {12},
    number = {7},
    pages = {1030--1051},
    year = {2024},
    publisher = {Routledge},
    doi = {10.1080/21670811.2024.2342468},
    URL = {https://doi.org/10.1080/21670811.2024.2342468},
    eprint = {https://doi.org/10.1080/21670811.2024.2342468}
}

@article{D’haeseleer13082025,
    author = {Stephanie D’haeseleer and Kristin Van Damme and Hannes Cools and Sarah Van Leuven and Tom Evens},
    title = {AI Divides in Newsrooms? How Journalists in the Low Countries Use and Perceive Generative AI},
    journal = {Journalism Practice},
    volume = {0},
    number = {0},
    pages = {1--28},
    year = {2025},
    publisher = {Routledge},
    doi = {10.1080/17512786.2025.2538120},
    URL = {https://doi.org/10.1080/17512786.2025.2538120},
    eprint = { https://doi.org/10.1080/17512786.2025.2538120}
}

@misc{anthropic2025claude4,
  author       = {{Anthropic}},
  title        = {Introducing Claude 4},
  year         = {2025},
  howpublished = {\url{https://www.anthropic.com/news/claude-4}},
  note         = {Accessed: 2025-08-22}
}

@misc{openai2025gpt5,
  author       = {{OpenAI}},
  title        = {Introducing GPT-5},
  year         = {2025},
  howpublished = {\url{https://openai.com/index/introducing-gpt-5/}},
  note         = {Published August 7, 2025; Accessed: 2025-08-22}
}

@inproceedings{10.1145/3630106.3659012,
author = {Orr, Will and Kang, Edward B.},
title = {AI as a Sport: On the Competitive Epistemologies of Benchmarking},
year = {2024},
isbn = {9798400704505},
publisher = {Association for Computing Machinery},
address = {New York, NY, USA},
url = {https://doi.org/10.1145/3630106.3659012},
doi = {10.1145/3630106.3659012},
booktitle = {Proceedings of the 2024 ACM Conference on Fairness, Accountability, and Transparency},
pages = {1875–1884},
numpages = {10},
location = {Rio de Janeiro, Brazil},
series = {FAccT '24}
}

@inproceedings{raji2021ai,
    title={{AI} and the Everything in the Whole Wide World Benchmark},
    author={Inioluwa Deborah Raji and Emily Denton and Emily M. Bender and Alex Hanna and Amandalynne Paullada},
    booktitle={Thirty-fifth Conference on Neural Information Processing Systems Datasets and Benchmarks Track (Round 2)},
    year={2021},
    url={https://openreview.net/forum?id=j6NxpQbREA1}
}

@misc{saxon2024benchmarksmicroscopesmodelmetrology,
      title={Benchmarks as Microscopes: A Call for Model Metrology}, 
      author={Michael Saxon and Ari Holtzman and Peter West and William Yang Wang and Naomi Saphra},
      year={2024},
      eprint={2407.16711},
      archivePrefix={arXiv},
      primaryClass={cs.SE},
      url={https://arxiv.org/abs/2407.16711}, 
}

@misc{liao2025rethinkingmodelevaluationnarrowing,
      title={Rethinking Model Evaluation as Narrowing the Socio-Technical Gap}, 
      author={Q. Vera Liao and Ziang Xiao},
      year={2025},
      eprint={2306.03100},
      archivePrefix={arXiv},
      primaryClass={cs.HC},
      url={https://arxiv.org/abs/2306.03100}, 
}

@misc{ethayarajh2021utilityeyeusercritique,
      title={Utility is in the Eye of the User: A Critique of NLP Leaderboards}, 
      author={Kawin Ethayarajh and Dan Jurafsky},
      year={2021},
      eprint={2009.13888},
      archivePrefix={arXiv},
      primaryClass={cs.CL},
      url={https://arxiv.org/abs/2009.13888}, 
}

@misc{reuel2024betterbenchassessingaibenchmarks,
      title={BetterBench: Assessing AI Benchmarks, Uncovering Issues, and Establishing Best Practices}, 
      author={Anka Reuel and Amelia Hardy and Chandler Smith and Max Lamparth and Malcolm Hardy and Mykel J. Kochenderfer},
      year={2024},
      eprint={2411.12990},
      archivePrefix={arXiv},
      primaryClass={cs.AI},
      url={https://arxiv.org/abs/2411.12990}, 
}

@misc{liu2024ecbdevidencecenteredbenchmarkdesign,
      title={ECBD: Evidence-Centered Benchmark Design for NLP}, 
      author={Yu Lu Liu and Su Lin Blodgett and Jackie Chi Kit Cheung and Q. Vera Liao and Alexandra Olteanu and Ziang Xiao},
      year={2024},
      eprint={2406.08723},
      archivePrefix={arXiv},
      primaryClass={cs.CL},
      url={https://arxiv.org/abs/2406.08723}, 
}

@inproceedings{10.1145/3630106.3658992,
author = {Suresh, Harini and Tseng, Emily and Young, Meg and Gray, Mary and Pierson, Emma and Levy, Karen},
title = {Participation in the age of foundation models},
year = {2024},
isbn = {9798400704505},
publisher = {Association for Computing Machinery},
address = {New York, NY, USA},
url = {https://doi.org/10.1145/3630106.3658992},
doi = {10.1145/3630106.3658992},
booktitle = {Proceedings of the 2024 ACM Conference on Fairness, Accountability, and Transparency},
pages = {1609–1621},
numpages = {13},
location = {Rio de Janeiro, Brazil},
series = {FAccT '24}
}

@article{cools2024uses-635, 
  year    = {2024}, 
  title   = {Uses of Generative {AI} in the Newsroom: Mapping Journalists’ Perceptions of Perils and Possibilities}, 
  author  = {Cools, Hannes and Diakopoulos, Nicholas}, 
  journal = {Journalism Practice}, 
  issn    = {1751-2786}, 
  doi     = {10.1080/17512786.2024.2394558}, 
  pages   = {1--19}, 
  number  = {ahead-of-print}, 
  volume  = {ahead-of-print}
}

@article{li2024newsbench-770, 
  year    = {2024}, 
  title   = {{NewsBench}: A Systematic Evaluation Framework for Assessing Editorial Capabilities of Large Language Models in Chinese Journalism}, 
  author  = {Li, Miao and Chen, Ming-Bin and Tang, Bo and {ShengbinHou}, {ShengbinHou} and Wang, Pengyu and Deng, Haiying and Li, Zhiyu and Xiong, Feiyu and Mao, Keming and Peng, Cheng and Luo, Yi}, 
  journal = {Proceedings of the 62nd Annual Meeting of the Association for Computational Linguistics (Volume 1: Long Papers)}, 
  doi     = {10.18653/v1/2024.acl-long.538}, 
  pages   = {9993--10014}
}

@article{nishal2024envisioning-891, 
  year    = {2024}, 
  title   = {Envisioning the Applications and Implications of Generative {AI} for News Media}, 
  author  = {Nishal, Sachita and Diakopoulos, Nicholas}, 
  journal = {{arXiv}}, 
  doi     = {10.48550/arxiv.2402.18835}, 
  eprint  = {2402.18835}
}

@article{tseng2025ownership-8b2, 
  year    = {2025}, 
  title   = {"Ownership, Not Just Happy Talk": Co-Designing a Participatory Large Language Model for Journalism}, 
  author  = {Tseng, Emily and Young, Meg and Quéré, Marianne Aubin Le and Rinehart, Aimee and Suresh, Harini}, 
  journal = {{arXiv}}, 
  eprint  = {2501.17299}
}

@article{diakopoulos2023leveraging-af4, 
  year    = {2023}, 
  title   = {Leveraging Professional Ethics for Responsible {AI}}, 
  author  = {Diakopoulos, Nicholas and Trattner, Christoph and Jannach, Dietmar and Meijer, Irene Costera and Motta, Enrico}, 
  journal = {Communications of the {ACM}}, 
  issn    = {0001-0782}, 
  doi     = {10.1145/3625252}
}

@misc{matton2024leakagecodegenerationevaluation,
      title={On Leakage of Code Generation Evaluation Datasets}, 
      author={Alexandre Matton and Tom Sherborne and Dennis Aumiller and Elena Tommasone and Milad Alizadeh and Jingyi He and Raymond Ma and Maxime Voisin and Ellen Gilsenan-McMahon and Matthias Gallé},
      year={2024},
      eprint={2407.07565},
      archivePrefix={arXiv},
      primaryClass={cs.CL},
      url={https://arxiv.org/abs/2407.07565}, 
}

@misc{chen2021evaluatinglargelanguagemodels,
      title={Evaluating Large Language Models Trained on Code}, 
      author={Mark Chen and Jerry Tworek and Heewoo Jun and Qiming Yuan and Henrique Ponde de Oliveira Pinto and Jared Kaplan and Harri Edwards and Yuri Burda and Nicholas Joseph and Greg Brockman and Alex Ray and Raul Puri and Gretchen Krueger and Michael Petrov and Heidy Khlaaf and Girish Sastry and Pamela Mishkin and Brooke Chan and Scott Gray and Nick Ryder and Mikhail Pavlov and Alethea Power and Lukasz Kaiser and Mohammad Bavarian and Clemens Winter and Philippe Tillet and Felipe Petroski Such and Dave Cummings and Matthias Plappert and Fotios Chantzis and Elizabeth Barnes and Ariel Herbert-Voss and William Hebgen Guss and Alex Nichol and Alex Paino and Nikolas Tezak and Jie Tang and Igor Babuschkin and Suchir Balaji and Shantanu Jain and William Saunders and Christopher Hesse and Andrew N. Carr and Jan Leike and Josh Achiam and Vedant Misra and Evan Morikawa and Alec Radford and Matthew Knight and Miles Brundage and Mira Murati and Katie Mayer and Peter Welinder and Bob McGrew and Dario Amodei and Sam McCandlish and Ilya Sutskever and Wojciech Zaremba},
      year={2021},
      eprint={2107.03374},
      archivePrefix={arXiv},
      primaryClass={cs.LG},
      url={https://arxiv.org/abs/2107.03374}, 
}

@misc{austin2021programsynthesislargelanguage,
      title={Program Synthesis with Large Language Models}, 
      author={Jacob Austin and Augustus Odena and Maxwell Nye and Maarten Bosma and Henryk Michalewski and David Dohan and Ellen Jiang and Carrie Cai and Michael Terry and Quoc Le and Charles Sutton},
      year={2021},
      eprint={2108.07732},
      archivePrefix={arXiv},
      primaryClass={cs.PL},
      url={https://arxiv.org/abs/2108.07732}, 
}

@inproceedings{yu-etal-2018-spider,
    title = "{S}pider: A Large-Scale Human-Labeled Dataset for Complex and Cross-Domain Semantic Parsing and Text-to-{SQL} Task",
    author = "Yu, Tao  and
      Zhang, Rui  and
      Yang, Kai  and
      Yasunaga, Michihiro  and
      Wang, Dongxu  and
      Li, Zifan  and
      Ma, James  and
      Li, Irene  and
      Yao, Qingning  and
      Roman, Shanelle  and
      Zhang, Zilin  and
      Radev, Dragomir",
    editor = "Riloff, Ellen  and
      Chiang, David  and
      Hockenmaier, Julia  and
      Tsujii, Jun{'}ichi",
    booktitle = "Proceedings of the 2018 Conference on Empirical Methods in Natural Language Processing",
    month = oct # "-" # nov,
    year = "2018",
    address = "Brussels, Belgium",
    publisher = "Association for Computational Linguistics",
    url = "https://aclanthology.org/D18-1425/",
    doi = "10.18653/v1/D18-1425",
    pages = "3911--3921"
}

@inproceedings{Bisk2019PIQARA,
  title={PIQA: Reasoning about Physical Commonsense in Natural Language},
  author={Yonatan Bisk and Rowan Zellers and Ronan Le Bras and Jianfeng Gao and Yejin Choi},
  booktitle={AAAI Conference on Artificial Intelligence},
  year={2019},
  url={https://api.semanticscholar.org/CorpusID:208290939}
}

@inproceedings{zellers-etal-2019-hellaswag,
    title = "{H}ella{S}wag: Can a Machine Really Finish Your Sentence?",
    author = "Zellers, Rowan  and
      Holtzman, Ari  and
      Bisk, Yonatan  and
      Farhadi, Ali  and
      Choi, Yejin",
    editor = "Korhonen, Anna  and
      Traum, David  and
      M{\`a}rquez, Llu{\'i}s",
    booktitle = "Proceedings of the 57th Annual Meeting of the Association for Computational Linguistics",
    month = jul,
    year = "2019",
    address = "Florence, Italy",
    publisher = "Association for Computational Linguistics",
    url = "https://aclanthology.org/P19-1472/",
    doi = "10.18653/v1/P19-1472",
    pages = "4791--4800"
}

@misc{sakaguchi2019winograndeadversarialwinogradschema,
      title={WinoGrande: An Adversarial Winograd Schema Challenge at Scale}, 
      author={Keisuke Sakaguchi and Ronan Le Bras and Chandra Bhagavatula and Yejin Choi},
      year={2019},
      eprint={1907.10641},
      archivePrefix={arXiv},
      primaryClass={cs.CL},
      url={https://arxiv.org/abs/1907.10641}, 
}

@misc{dua2019dropreadingcomprehensionbenchmark,
      title={DROP: A Reading Comprehension Benchmark Requiring Discrete Reasoning Over Paragraphs}, 
      author={Dheeru Dua and Yizhong Wang and Pradeep Dasigi and Gabriel Stanovsky and Sameer Singh and Matt Gardner},
      year={2019},
      eprint={1903.00161},
      archivePrefix={arXiv},
      primaryClass={cs.CL},
      url={https://arxiv.org/abs/1903.00161}, 
}

@inproceedings{rajpurkar-etal-2018-know,
    title = "Know What You Don{'}t Know: Unanswerable Questions for {SQ}u{AD}",
    author = "Rajpurkar, Pranav  and
      Jia, Robin  and
      Liang, Percy",
    editor = "Gurevych, Iryna  and
      Miyao, Yusuke",
    booktitle = "Proceedings of the 56th Annual Meeting of the Association for Computational Linguistics (Volume 2: Short Papers)",
    month = jul,
    year = "2018",
    address = "Melbourne, Australia",
    publisher = "Association for Computational Linguistics",
    url = "https://aclanthology.org/P18-2124/",
    doi = "10.18653/v1/P18-2124",
    pages = "784--789"
}

@inproceedings{lai-etal-2017-race,
    title = "{RACE}: Large-scale {R}e{A}ding Comprehension Dataset From Examinations",
    author = "Lai, Guokun  and
      Xie, Qizhe  and
      Liu, Hanxiao  and
      Yang, Yiming  and
      Hovy, Eduard",
    editor = "Palmer, Martha  and
      Hwa, Rebecca  and
      Riedel, Sebastian",
    booktitle = "Proceedings of the 2017 Conference on Empirical Methods in Natural Language Processing",
    month = sep,
    year = "2017",
    address = "Copenhagen, Denmark",
    publisher = "Association for Computational Linguistics",
    url = "https://aclanthology.org/D17-1082/",
    doi = "10.18653/v1/D17-1082",
    pages = "785--794"
}

@misc{cobbe2021trainingverifierssolvemath,
      title={Training Verifiers to Solve Math Word Problems}, 
      author={Karl Cobbe and Vineet Kosaraju and Mohammad Bavarian and Mark Chen and Heewoo Jun and Lukasz Kaiser and Matthias Plappert and Jerry Tworek and Jacob Hilton and Reiichiro Nakano and Christopher Hesse and John Schulman},
      year={2021},
      eprint={2110.14168},
      archivePrefix={arXiv},
      primaryClass={cs.LG},
      url={https://arxiv.org/abs/2110.14168}, 
}

@misc{liang2023holisticevaluationlanguagemodels,
      title={Holistic Evaluation of Language Models}, 
      author={Percy Liang and Rishi Bommasani and Tony Lee and Dimitris Tsipras and Dilara Soylu and Michihiro Yasunaga and Yian Zhang and Deepak Narayanan and Yuhuai Wu and Ananya Kumar and Benjamin Newman and Binhang Yuan and Bobby Yan and Ce Zhang and Christian Cosgrove and Christopher D. Manning and Christopher Ré and Diana Acosta-Navas and Drew A. Hudson and Eric Zelikman and Esin Durmus and Faisal Ladhak and Frieda Rong and Hongyu Ren and Huaxiu Yao and Jue Wang and Keshav Santhanam and Laurel Orr and Lucia Zheng and Mert Yuksekgonul and Mirac Suzgun and Nathan Kim and Neel Guha and Niladri Chatterji and Omar Khattab and Peter Henderson and Qian Huang and Ryan Chi and Sang Michael Xie and Shibani Santurkar and Surya Ganguli and Tatsunori Hashimoto and Thomas Icard and Tianyi Zhang and Vishrav Chaudhary and William Wang and Xuechen Li and Yifan Mai and Yuhui Zhang and Yuta Koreeda},
      year={2023},
      eprint={2211.09110},
      archivePrefix={arXiv},
      primaryClass={cs.CL},
      url={https://arxiv.org/abs/2211.09110}, 
}

@inproceedings{hendrycks2021measuring,
    title={Measuring Mathematical Problem Solving With the {MATH} Dataset},
    author={Dan Hendrycks and Collin Burns and Saurav Kadavath and Akul Arora and Steven Basart and Eric Tang and Dawn Song and Jacob Steinhardt},
    booktitle={Thirty-fifth Conference on Neural Information Processing Systems Datasets and Benchmarks Track (Round 2)},
    year={2021},
    url={https://openreview.net/forum?id=7Bywt2mQsCe}
}

@misc{srivastava2023imitationgamequantifyingextrapolating,
      title={Beyond the Imitation Game: Quantifying and extrapolating the capabilities of language models}, 
      author={Aarohi Srivastava and Abhinav Rastogi and Abhishek Rao and Abu Awal Md Shoeb and Abubakar Abid and Adam Fisch and Adam R. Brown and Adam Santoro and Aditya Gupta and Adrià Garriga-Alonso and Agnieszka Kluska and Aitor Lewkowycz and Akshat Agarwal and Alethea Power and Alex Ray and Alex Warstadt and Alexander W. Kocurek and Ali Safaya and Ali Tazarv and Alice Xiang and Alicia Parrish and Allen Nie and Aman Hussain and Amanda Askell and Amanda Dsouza and Ambrose Slone and Ameet Rahane and Anantharaman S. Iyer and Anders Andreassen and Andrea Madotto and Andrea Santilli and Andreas Stuhlmüller and Andrew Dai and Andrew La and Andrew Lampinen and Andy Zou and Angela Jiang and Angelica Chen and Anh Vuong and Animesh Gupta and Anna Gottardi and Antonio Norelli and Anu Venkatesh and Arash Gholamidavoodi and Arfa Tabassum and Arul Menezes and Arun Kirubarajan and Asher Mullokandov and Ashish Sabharwal and Austin Herrick and Avia Efrat and Aykut Erdem and Ayla Karakaş and B. Ryan Roberts and Bao Sheng Loe and Barret Zoph and Bartłomiej Bojanowski and Batuhan Özyurt and Behnam Hedayatnia and Behnam Neyshabur and Benjamin Inden and Benno Stein and Berk Ekmekci and Bill Yuchen Lin and Blake Howald and Bryan Orinion and Cameron Diao and Cameron Dour and Catherine Stinson and Cedrick Argueta and César Ferri Ramírez and Chandan Singh and Charles Rathkopf and Chenlin Meng and Chitta Baral and Chiyu Wu and Chris Callison-Burch and Chris Waites and Christian Voigt and Christopher D. Manning and Christopher Potts and Cindy Ramirez and Clara E. Rivera and Clemencia Siro and Colin Raffel and Courtney Ashcraft and Cristina Garbacea and Damien Sileo and Dan Garrette and Dan Hendrycks and Dan Kilman and Dan Roth and Daniel Freeman and Daniel Khashabi and Daniel Levy and Daniel Moseguí González and Danielle Perszyk and Danny Hernandez and Danqi Chen and Daphne Ippolito and Dar Gilboa and David Dohan and David Drakard and David Jurgens and Debajyoti Datta and Deep Ganguli and Denis Emelin and Denis Kleyko and Deniz Yuret and Derek Chen and Derek Tam and Dieuwke Hupkes and Diganta Misra and Dilyar Buzan and Dimitri Coelho Mollo and Diyi Yang and Dong-Ho Lee and Dylan Schrader and Ekaterina Shutova and Ekin Dogus Cubuk and Elad Segal and Eleanor Hagerman and Elizabeth Barnes and Elizabeth Donoway and Ellie Pavlick and Emanuele Rodola and Emma Lam and Eric Chu and Eric Tang and Erkut Erdem and Ernie Chang and Ethan A. Chi and Ethan Dyer and Ethan Jerzak and Ethan Kim and Eunice Engefu Manyasi and Evgenii Zheltonozhskii and Fanyue Xia and Fatemeh Siar and Fernando Martínez-Plumed and Francesca Happé and Francois Chollet and Frieda Rong and Gaurav Mishra and Genta Indra Winata and Gerard de Melo and Germán Kruszewski and Giambattista Parascandolo and Giorgio Mariani and Gloria Wang and Gonzalo Jaimovitch-López and Gregor Betz and Guy Gur-Ari and Hana Galijasevic and Hannah Kim and Hannah Rashkin and Hannaneh Hajishirzi and Harsh Mehta and Hayden Bogar and Henry Shevlin and Hinrich Schütze and Hiromu Yakura and Hongming Zhang and Hugh Mee Wong and Ian Ng and Isaac Noble and Jaap Jumelet and Jack Geissinger and Jackson Kernion and Jacob Hilton and Jaehoon Lee and Jaime Fernández Fisac and James B. Simon and James Koppel and James Zheng and James Zou and Jan Kocoń and Jana Thompson and Janelle Wingfield and Jared Kaplan and Jarema Radom and Jascha Sohl-Dickstein and Jason Phang and Jason Wei and Jason Yosinski and Jekaterina Novikova and Jelle Bosscher and Jennifer Marsh and Jeremy Kim and Jeroen Taal and Jesse Engel and Jesujoba Alabi and Jiacheng Xu and Jiaming Song and Jillian Tang and Joan Waweru and John Burden and John Miller and John U. Balis and Jonathan Batchelder and Jonathan Berant and Jörg Frohberg and Jos Rozen and Jose Hernandez-Orallo and Joseph Boudeman and Joseph Guerr and Joseph Jones and Joshua B. Tenenbaum and Joshua S. Rule and Joyce Chua and Kamil Kanclerz and Karen Livescu and Karl Krauth and Karthik Gopalakrishnan and Katerina Ignatyeva and Katja Markert and Kaustubh D. Dhole and Kevin Gimpel and Kevin Omondi and Kory Mathewson and Kristen Chiafullo and Ksenia Shkaruta and Kumar Shridhar and Kyle McDonell and Kyle Richardson and Laria Reynolds and Leo Gao and Li Zhang and Liam Dugan and Lianhui Qin and Lidia Contreras-Ochando and Louis-Philippe Morency and Luca Moschella and Lucas Lam and Lucy Noble and Ludwig Schmidt and Luheng He and Luis Oliveros Colón and Luke Metz and Lütfi Kerem Şenel and Maarten Bosma and Maarten Sap and Maartje ter Hoeve and Maheen Farooqi and Manaal Faruqui and Mantas Mazeika and Marco Baturan and Marco Marelli and Marco Maru and Maria Jose Ramírez Quintana and Marie Tolkiehn and Mario Giulianelli and Martha Lewis and Martin Potthast and Matthew L. Leavitt and Matthias Hagen and Mátyás Schubert and Medina Orduna Baitemirova and Melody Arnaud and Melvin McElrath and Michael A. Yee and Michael Cohen and Michael Gu and Michael Ivanitskiy and Michael Starritt and Michael Strube and Michał Swędrowski and Michele Bevilacqua and Michihiro Yasunaga and Mihir Kale and Mike Cain and Mimee Xu and Mirac Suzgun and Mitch Walker and Mo Tiwari and Mohit Bansal and Moin Aminnaseri and Mor Geva and Mozhdeh Gheini and Mukund Varma T and Nanyun Peng and Nathan A. Chi and Nayeon Lee and Neta Gur-Ari Krakover and Nicholas Cameron and Nicholas Roberts and Nick Doiron and Nicole Martinez and Nikita Nangia and Niklas Deckers and Niklas Muennighoff and Nitish Shirish Keskar and Niveditha S. Iyer and Noah Constant and Noah Fiedel and Nuan Wen and Oliver Zhang and Omar Agha and Omar Elbaghdadi and Omer Levy and Owain Evans and Pablo Antonio Moreno Casares and Parth Doshi and Pascale Fung and Paul Pu Liang and Paul Vicol and Pegah Alipoormolabashi and Peiyuan Liao and Percy Liang and Peter Chang and Peter Eckersley and Phu Mon Htut and Pinyu Hwang and Piotr Miłkowski and Piyush Patil and Pouya Pezeshkpour and Priti Oli and Qiaozhu Mei and Qing Lyu and Qinlang Chen and Rabin Banjade and Rachel Etta Rudolph and Raefer Gabriel and Rahel Habacker and Ramon Risco and Raphaël Millière and Rhythm Garg and Richard Barnes and Rif A. Saurous and Riku Arakawa and Robbe Raymaekers and Robert Frank and Rohan Sikand and Roman Novak and Roman Sitelew and Ronan LeBras and Rosanne Liu and Rowan Jacobs and Rui Zhang and Ruslan Salakhutdinov and Ryan Chi and Ryan Lee and Ryan Stovall and Ryan Teehan and Rylan Yang and Sahib Singh and Saif M. Mohammad and Sajant Anand and Sam Dillavou and Sam Shleifer and Sam Wiseman and Samuel Gruetter and Samuel R. Bowman and Samuel S. Schoenholz and Sanghyun Han and Sanjeev Kwatra and Sarah A. Rous and Sarik Ghazarian and Sayan Ghosh and Sean Casey and Sebastian Bischoff and Sebastian Gehrmann and Sebastian Schuster and Sepideh Sadeghi and Shadi Hamdan and Sharon Zhou and Shashank Srivastava and Sherry Shi and Shikhar Singh and Shima Asaadi and Shixiang Shane Gu and Shubh Pachchigar and Shubham Toshniwal and Shyam Upadhyay and Shyamolima and Debnath and Siamak Shakeri and Simon Thormeyer and Simone Melzi and Siva Reddy and Sneha Priscilla Makini and Soo-Hwan Lee and Spencer Torene and Sriharsha Hatwar and Stanislas Dehaene and Stefan Divic and Stefano Ermon and Stella Biderman and Stephanie Lin and Stephen Prasad and Steven T. Piantadosi and Stuart M. Shieber and Summer Misherghi and Svetlana Kiritchenko and Swaroop Mishra and Tal Linzen and Tal Schuster and Tao Li and Tao Yu and Tariq Ali and Tatsu Hashimoto and Te-Lin Wu and Théo Desbordes and Theodore Rothschild and Thomas Phan and Tianle Wang and Tiberius Nkinyili and Timo Schick and Timofei Kornev and Titus Tunduny and Tobias Gerstenberg and Trenton Chang and Trishala Neeraj and Tushar Khot and Tyler Shultz and Uri Shaham and Vedant Misra and Vera Demberg and Victoria Nyamai and Vikas Raunak and Vinay Ramasesh and Vinay Uday Prabhu and Vishakh Padmakumar and Vivek Srikumar and William Fedus and William Saunders and William Zhang and Wout Vossen and Xiang Ren and Xiaoyu Tong and Xinran Zhao and Xinyi Wu and Xudong Shen and Yadollah Yaghoobzadeh and Yair Lakretz and Yangqiu Song and Yasaman Bahri and Yejin Choi and Yichi Yang and Yiding Hao and Yifu Chen and Yonatan Belinkov and Yu Hou and Yufang Hou and Yuntao Bai and Zachary Seid and Zhuoye Zhao and Zijian Wang and Zijie J. Wang and Zirui Wang and Ziyi Wu},
      year={2023},
      eprint={2206.04615},
      archivePrefix={arXiv},
      primaryClass={cs.CL},
      url={https://arxiv.org/abs/2206.04615}, 
}

@misc{zhong2023agievalhumancentricbenchmarkevaluating,
      title={AGIEval: A Human-Centric Benchmark for Evaluating Foundation Models}, 
      author={Wanjun Zhong and Ruixiang Cui and Yiduo Guo and Yaobo Liang and Shuai Lu and Yanlin Wang and Amin Saied and Weizhu Chen and Nan Duan},
      year={2023},
      eprint={2304.06364},
      archivePrefix={arXiv},
      primaryClass={cs.CL},
      url={https://arxiv.org/abs/2304.06364}, 
}

@article{10.1145/3615355,
    author = {Davis, Ernest},
    title = {Benchmarks for Automated Commonsense Reasoning: A Survey},
    year = {2023},
    issue_date = {April 2024},
    publisher = {Association for Computing Machinery},
    address = {New York, NY, USA},
    volume = {56},
    number = {4},
    issn = {0360-0300},
    url = {https://doi.org/10.1145/3615355},
    doi = {10.1145/3615355},
    journal = {ACM Comput. Surv.},
    month = oct,
    articleno = {81},
    numpages = {41}
}

@misc{biderman2024lessonstrenchesreproducibleevaluation,
      title={Lessons from the Trenches on Reproducible Evaluation of Language Models}, 
      author={Stella Biderman and Hailey Schoelkopf and Lintang Sutawika and Leo Gao and Jonathan Tow and Baber Abbasi and Alham Fikri Aji and Pawan Sasanka Ammanamanchi and Sidney Black and Jordan Clive and Anthony DiPofi and Julen Etxaniz and Benjamin Fattori and Jessica Zosa Forde and Charles Foster and Jeffrey Hsu and Mimansa Jaiswal and Wilson Y. Lee and Haonan Li and Charles Lovering and Niklas Muennighoff and Ellie Pavlick and Jason Phang and Aviya Skowron and Samson Tan and Xiangru Tang and Kevin A. Wang and Genta Indra Winata and François Yvon and Andy Zou},
      year={2024},
      eprint={2405.14782},
      archivePrefix={arXiv},
      primaryClass={cs.CL},
      url={https://arxiv.org/abs/2405.14782}, 
}

@misc{guha2023legalbenchcollaborativelybuiltbenchmark,
      title={LegalBench: A Collaboratively Built Benchmark for Measuring Legal Reasoning in Large Language Models}, 
      author={Neel Guha and Julian Nyarko and Daniel E. Ho and Christopher Ré and Adam Chilton and Aditya Narayana and Alex Chohlas-Wood and Austin Peters and Brandon Waldon and Daniel N. Rockmore and Diego Zambrano and Dmitry Talisman and Enam Hoque and Faiz Surani and Frank Fagan and Galit Sarfaty and Gregory M. Dickinson and Haggai Porat and Jason Hegland and Jessica Wu and Joe Nudell and Joel Niklaus and John Nay and Jonathan H. Choi and Kevin Tobia and Margaret Hagan and Megan Ma and Michael Livermore and Nikon Rasumov-Rahe and Nils Holzenberger and Noam Kolt and Peter Henderson and Sean Rehaag and Sharad Goel and Shang Gao and Spencer Williams and Sunny Gandhi and Tom Zur and Varun Iyer and Zehua Li},
      year={2023},
      eprint={2308.11462},
      archivePrefix={arXiv},
      primaryClass={cs.CL},
      url={https://arxiv.org/abs/2308.11462}, 
}

@misc{rein2023gpqagraduatelevelgoogleproofqa,
      title={GPQA: A Graduate-Level Google-Proof Q\&A Benchmark}, 
      author={David Rein and Betty Li Hou and Asa Cooper Stickland and Jackson Petty and Richard Yuanzhe Pang and Julien Dirani and Julian Michael and Samuel R. Bowman},
      year={2023},
      eprint={2311.12022},
      archivePrefix={arXiv},
      primaryClass={cs.AI},
      url={https://arxiv.org/abs/2311.12022}, 
}

@article{reddy2021evaluation,
  author    = {Reddy, Sandeep and Rogers, Wendy and Makinen, Ville-Petteri and Coiera, Enrico and Brown, Pieta and Wenzel, Markus and Weicken, Eva and Ansari, Saba and Mathur, Piyush and Casey, Aaron and Kelly, Blair},
  title     = {Evaluation Framework to Guide Implementation of AI Systems into Healthcare Settings},
  journal   = {BMJ Health \& Care Informatics},
  volume    = {28},
  number    = {1},
  year      = {2021},
  pages     = {e100444},
  doi       = {10.1136/bmjhci-2021-100444},
  url       = {https://doi.org/10.1136/bmjhci-2021-100444}
}

@inproceedings{jimenez2024swebench,
    title={{SWE}-bench: Can Language Models Resolve Real-world Github Issues?},
    author={Carlos E Jimenez and John Yang and Alexander Wettig and Shunyu Yao and Kexin Pei and Ofir Press and Karthik R Narasimhan},
    booktitle={The Twelfth International Conference on Learning Representations},
    year={2024},
    url={https://openreview.net/forum?id=VTF8yNQM66}
}

@misc{yao2024taubenchbenchmarktoolagentuserinteraction,
      title={$\tau$-bench: A Benchmark for Tool-Agent-User Interaction in Real-World Domains}, 
      author={Shunyu Yao and Noah Shinn and Pedram Razavi and Karthik Narasimhan},
      year={2024},
      eprint={2406.12045},
      archivePrefix={arXiv},
      primaryClass={cs.AI},
      url={https://arxiv.org/abs/2406.12045}, 
}

@inproceedings{OHara2015ComputationalNF,
  title={Computational Notebooks for AI Education},
  author={Keith J. O'Hara and Douglas S. Blank and James B. Marshall},
  booktitle={The Florida AI Research Society},
  year={2015},
  url={https://api.semanticscholar.org/CorpusID:1772160}
}

@article{Deuze2005if, 
  year     = {2005}, 
  title    = {What is journalism?: Professional identity and ideology of journalists reconsidered}, 
  author   = {Deuze, M}, 
  journal  = {Journalism}, 
  doi      = {10.1177/1464884905056815}, 
  pages    = {442 -- 464}, 
  number   = {4}, 
  volume   = {6}, 
  language = {English}, 
  month    = {11}
}

@article{Hanitzsch2007it, 
  year     = {2007}, 
  title    = {Deconstructing Journalism Culture: Toward a Universal Theory}, 
  author   = {Hanitzsch, Thomas}, 
  journal  = {Communication theory}, 
  doi      = {10.1111/j.1468-2885.2007.00303.x}, 
  pages    = {367 -- 385}, 
  number   = {4}, 
  volume   = {17}, 
  language = {English}, 
  month    = {11}
}

@inproceedings{10.1145/3708359.3712152,
author = {Hardy, Amelia and Reuel, Anka and Jafari Meimandi, Kiana and Soder, Lisa and Griffith, Allie and Asmar, Dylan M and Koyejo, Sanmi and Bernstein, Michael S. and Kochenderfer, Mykel John},
title = {More than Marketing? On the Information Value of AI Benchmarks for Practitioners},
year = {2025},
isbn = {9798400713064},
publisher = {Association for Computing Machinery},
address = {New York, NY, USA},
url = {https://doi.org/10.1145/3708359.3712152},
doi = {10.1145/3708359.3712152},
booktitle = {Proceedings of the 30th International Conference on Intelligent User Interfaces},
pages = {1032–1047},
numpages = {16},
location = {
},
series = {IUI '25}
}

@misc{schwartz2025realitychecknewevaluation,
      title={Reality Check: A New Evaluation Ecosystem Is Necessary to Understand AI's Real World Effects}, 
      author={Reva Schwartz and Rumman Chowdhury and Akash Kundu and Heather Frase and Marzieh Fadaee and Tom David and Gabriella Waters and Afaf Taik and Morgan Briggs and Patrick Hall and Shomik Jain and Kyra Yee and Spencer Thomas and Sundeep Bhandari and Paul Duncan and Andrew Thompson and Maya Carlyle and Qinghua Lu and Matthew Holmes and Theodora Skeadas},
      year={2025},
      eprint={2505.18893},
      archivePrefix={arXiv},
      primaryClass={cs.CY},
      url={https://arxiv.org/abs/2505.18893}, 
}

@article{Young20102024,
    author = {Mary Lynn Young and Alfred Hermida},
    title = {People, Power, Platforms and the Business of Journalism},
    journal = {Digital Journalism},
    volume = {12},
    number = {9},
    pages = {1250--1260},
    year = {2024},
    publisher = {Routledge},
    doi = {10.1080/21670811.2023.2273523},
    URL = {https://doi.org/10.1080/21670811.2023.2273523},
    eprint = { https://doi.org/10.1080/21670811.2023.2273523}
}

@article{10.1162/DESI,
    author = {Steen, Marc},
    title = {Co-Design as a Process of Joint Inquiry and Imagination},
    journal = {Design Issues},
    volume = {29},
    number = {2},
    pages = {16-28},
    year = {2013},
    month = {04},
    issn = {0747-9360},
    doi = {10.1162/DESI\_a\_00207},
    url = {https://doi.org/10.1162/DESI\_a\_00207},
    eprint = {https://direct.mit.edu/desi/article-pdf/29/2/16/1715163/desi\_a\_00207.pdf},
}

@article{braun2006using,
  title={Using thematic analysis in psychology},
  author={Braun, Virginia and Clarke, Victoria},
  journal={Qualitative research in psychology},
  volume={3},
  number={2},
  pages={77--101},
  year={2006},
  publisher={Taylor \& Francis}
}

@article{doi:10.1177/16094069241289278,
    author = {Colleen Cheek and Elizabeth Austin and Lieke Richardson and Luke Testa and Natalia Ransolin and Emilie Francis-Auton and Mariam Safi and Margaret Murphy and Aaron De Los Santos and Matthew Vukasovic and Robyn Clay-Williams},
    title ={Non-Participant Observations in Experience-Based Codesign: An example using a Case Study Research approach to explore Emergency Department Care},
    journal = {International Journal of Qualitative Methods},
    volume = {23},
    number = {},
    pages = {16094069241289278},
    year = {2024},
    doi = {10.1177/16094069241289278},
    URL ={https://doi.org/10.1177/16094069241289278},                     
    eprint = {https://doi.org/10.1177/16094069241289278}
}

@article{doi:10.1177/14648849241299775,
author = {Sora Park and Caroline Fisher and Edson TandocJr and Uwe Dulleck and Shengnan Pinker Yao and William Lukamto},
title ={The relationship between news trust, mistrust and audience disengagement},

journal = {Journalism},
volume = {0},
number = {0},
pages = {14648849241299775},
year = {0},
doi = {10.1177/14648849241299775},

URL = { 
    
        https://doi.org/10.1177/14648849241299775
    
    

},
eprint = { 
    
        https://doi.org/10.1177/14648849241299775
    
    

}
}

@article{campbell2020purposive,
  author  = {Campbell, Suzanne and Greenwood, Mike and Prior, Sarah and Shearer, Tom and Walkem, Kathryn and Young, Sally and Bywaters, David and Walker, Kate},
  title   = {Purposive Sampling: Complex or Simple? Research Case Examples},
  journal = {Journal of Research in Nursing},
  volume  = {25},
  number  = {8},
  pages   = {652--661},
  year    = {2020},
  doi     = {10.1177/1744987120927206},
  url     = {https://doi.org/10.1177/1744987120927206}
}

@Inbook{Ciesielska2018,
author="Ciesielska, Malgorzata
and Bostr{\"o}m, Katarzyna W.
and {\"O}hlander, Magnus",
editor="Ciesielska, Malgorzata
and Jemielniak, Dariusz",
title="Observation Methods",
bookTitle="Qualitative Methodologies in Organization Studies: Volume II: Methods and Possibilities",
year="2018",
publisher="Springer International Publishing",
address="Cham",
pages="33--52",
isbn="978-3-319-65442-3",
doi="10.1007/978-3-319-65442-3_2",
url="https://doi.org/10.1007/978-3-319-65442-3_2"
}

@misc{patwardhan2025gdpvalevaluatingaimodel,
      title={GDPval: Evaluating AI Model Performance on Real-World Economically Valuable Tasks}, 
      author={Tejal Patwardhan and Rachel Dias and Elizabeth Proehl and Grace Kim and Michele Wang and Olivia Watkins and Simón Posada Fishman and Marwan Aljubeh and Phoebe Thacker and Laurance Fauconnet and Natalie S. Kim and Patrick Chao and Samuel Miserendino and Gildas Chabot and David Li and Michael Sharman and Alexandra Barr and Amelia Glaese and Jerry Tworek},
      year={2025},
      eprint={2510.04374},
      archivePrefix={arXiv},
      primaryClass={cs.LG},
      url={https://arxiv.org/abs/2510.04374}, 
}

@article{10.63744@svxDtDD45mvw,
  title = {The Learnability Hierarchy of News Values: What Makes Some Journalistic Concepts Harder to Classify?},
  author = {Elisabeth Muth Andersen},
  year = {2025},
  journal = {Anthology of Computers and the Humanities},
  volume = {3},
  pages = {367--381},
  editor = {Taylor Arnold, Margherita Fantoli, and Ruben Ros},
  doi = {10.63744/svxDtDD45mvw}
}

@article{Harcup2016bn, 
  year     = {2016}, 
  title    = {What is news? News values revisited (again)}, 
  author   = {Harcup, Tony and O'Neill, Deirdre}, 
  journal  = {Journalism Studies}, 
  doi      = {10.1080/1461670x.2016.1150193}, 
  pages    = {1 -- 19}, 
  number   = {1}, 
  volume   = {23}, 
  language = {English}, 
  month    = {03}
}

@article{friedman2017survey-119, 
  year    = {2017}, 
  title   = {A Survey of Value Sensitive Design Methods}, 
  author  = {Friedman, Batya and Hendry, David G. and Borning, Alan}, 
  journal = {Foundations and Trends® in Human–Computer Interaction}, 
  issn    = {1551-3955}, 
  doi     = {10.1561/1100000015}, 
  pages   = {63--125}, 
  number  = {2}, 
  volume  = {11}
}

@misc{deepseekai2025deepseekv32pushingfrontieropen,
      title={DeepSeek-V3.2: Pushing the Frontier of Open Large Language Models}, 
      author={DeepSeek-AI and Aixin Liu and Aoxue Mei and Bangcai Lin and Bing Xue and Bingxuan Wang and Bingzheng Xu and Bochao Wu and Bowei Zhang and Chaofan Lin and Chen Dong and Chengda Lu and Chenggang Zhao and Chengqi Deng and Chenhao Xu and Chong Ruan and Damai Dai and Daya Guo and Dejian Yang and Deli Chen and Erhang Li and Fangqi Zhou and Fangyun Lin and Fucong Dai and Guangbo Hao and Guanting Chen and Guowei Li and H. Zhang and Hanwei Xu and Hao Li and Haofen Liang and Haoran Wei and Haowei Zhang and Haowen Luo and Haozhe Ji and Honghui Ding and Hongxuan Tang and Huanqi Cao and Huazuo Gao and Hui Qu and Hui Zeng and Jialiang Huang and Jiashi Li and Jiaxin Xu and Jiewen Hu and Jingchang Chen and Jingting Xiang and Jingyang Yuan and Jingyuan Cheng and Jinhua Zhu and Jun Ran and Junguang Jiang and Junjie Qiu and Junlong Li and Junxiao Song and Kai Dong and Kaige Gao and Kang Guan and Kexin Huang and Kexing Zhou and Kezhao Huang and Kuai Yu and Lean Wang and Lecong Zhang and Lei Wang and Liang Zhao and Liangsheng Yin and Lihua Guo and Lingxiao Luo and Linwang Ma and Litong Wang and Liyue Zhang and M. S. Di and M. Y Xu and Mingchuan Zhang and Minghua Zhang and Minghui Tang and Mingxu Zhou and Panpan Huang and Peixin Cong and Peiyi Wang and Qiancheng Wang and Qihao Zhu and Qingyang Li and Qinyu Chen and Qiushi Du and Ruiling Xu and Ruiqi Ge and Ruisong Zhang and Ruizhe Pan and Runji Wang and Runqiu Yin and Runxin Xu and Ruomeng Shen and Ruoyu Zhang and S. H. Liu and Shanghao Lu and Shangyan Zhou and Shanhuang Chen and Shaofei Cai and Shaoyuan Chen and Shengding Hu and Shengyu Liu and Shiqiang Hu and Shirong Ma and Shiyu Wang and Shuiping Yu and Shunfeng Zhou and Shuting Pan and Songyang Zhou and Tao Ni and Tao Yun and Tian Pei and Tian Ye and Tianyuan Yue and Wangding Zeng and Wen Liu and Wenfeng Liang and Wenjie Pang and Wenjing Luo and Wenjun Gao and Wentao Zhang and Xi Gao and Xiangwen Wang and Xiao Bi and Xiaodong Liu and Xiaohan Wang and Xiaokang Chen and Xiaokang Zhang and Xiaotao Nie and Xin Cheng and Xin Liu and Xin Xie and Xingchao Liu and Xingkai Yu and Xingyou Li and Xinyu Yang and Xinyuan Li and Xu Chen and Xuecheng Su and Xuehai Pan and Xuheng Lin and Xuwei Fu and Y. Q. Wang and Yang Zhang and Yanhong Xu and Yanru Ma and Yao Li and Yao Li and Yao Zhao and Yaofeng Sun and Yaohui Wang and Yi Qian and Yi Yu and Yichao Zhang and Yifan Ding and Yifan Shi and Yiliang Xiong and Ying He and Ying Zhou and Yinmin Zhong and Yishi Piao and Yisong Wang and Yixiao Chen and Yixuan Tan and Yixuan Wei and Yiyang Ma and Yiyuan Liu and Yonglun Yang and Yongqiang Guo and Yongtong Wu and Yu Wu and Yuan Cheng and Yuan Ou and Yuanfan Xu and Yuduan Wang and Yue Gong and Yuhan Wu and Yuheng Zou and Yukun Li and Yunfan Xiong and Yuxiang Luo and Yuxiang You and Yuxuan Liu and Yuyang Zhou and Z. F. Wu and Z. Z. Ren and Zehua Zhao and Zehui Ren and Zhangli Sha and Zhe Fu and Zhean Xu and Zhenda Xie and Zhengyan Zhang and Zhewen Hao and Zhibin Gou and Zhicheng Ma and Zhigang Yan and Zhihong Shao and Zhixian Huang and Zhiyu Wu and Zhuoshu Li and Zhuping Zhang and Zian Xu and Zihao Wang and Zihui Gu and Zijia Zhu and Zilin Li and Zipeng Zhang and Ziwei Xie and Ziyi Gao and Zizheng Pan and Zongqing Yao and Bei Feng and Hui Li and J. L. Cai and Jiaqi Ni and Lei Xu and Meng Li and Ning Tian and R. J. Chen and R. L. Jin and S. S. Li and Shuang Zhou and Tianyu Sun and X. Q. Li and Xiangyue Jin and Xiaojin Shen and Xiaosha Chen and Xinnan Song and Xinyi Zhou and Y. X. Zhu and Yanping Huang and Yaohui Li and Yi Zheng and Yuchen Zhu and Yunxian Ma and Zhen Huang and Zhipeng Xu and Zhongyu Zhang and Dongjie Ji and Jian Liang and Jianzhong Guo and Jin Chen and Leyi Xia and Miaojun Wang and Mingming Li and Peng Zhang and Ruyi Chen and Shangmian Sun and Shaoqing Wu and Shengfeng Ye and T. Wang and W. L. Xiao and Wei An and Xianzu Wang and Xiaowen Sun and Xiaoxiang Wang and Ying Tang and Yukun Zha and Zekai Zhang and Zhe Ju and Zhen Zhang and Zihua Qu},
      year={2025},
      eprint={2512.02556},
      archivePrefix={arXiv},
      primaryClass={cs.CL},
      url={https://arxiv.org/abs/2512.02556}, 
}

@misc{singh2025openaigpt5card,
      title={OpenAI GPT-5 System Card}, 
      author={Aaditya Singh and Adam Fry and Adam Perelman and Adam Tart and Adi Ganesh and Ahmed El-Kishky and Aidan McLaughlin and Aiden Low and AJ Ostrow and Akhila Ananthram and Akshay Nathan and Alan Luo and Alec Helyar and Aleksander Madry and Aleksandr Efremov and Aleksandra Spyra and Alex Baker-Whitcomb and Alex Beutel and Alex Karpenko and Alex Makelov and Alex Neitz and Alex Wei and Alexandra Barr and Alexandre Kirchmeyer and Alexey Ivanov and Alexi Christakis and Alistair Gillespie and Allison Tam and Ally Bennett and Alvin Wan and Alyssa Huang and Amy McDonald Sandjideh and Amy Yang and Ananya Kumar and Andre Saraiva and Andrea Vallone and Andrei Gheorghe and Andres Garcia Garcia and Andrew Braunstein and Andrew Liu and Andrew Schmidt and Andrey Mereskin and Andrey Mishchenko and Andy Applebaum and Andy Rogerson and Ann Rajan and Annie Wei and Anoop Kotha and Anubha Srivastava and Anushree Agrawal and Arun Vijayvergiya and Ashley Tyra and Ashvin Nair and Avi Nayak and Ben Eggers and Bessie Ji and Beth Hoover and Bill Chen and Blair Chen and Boaz Barak and Borys Minaiev and Botao Hao and Bowen Baker and Brad Lightcap and Brandon McKinzie and Brandon Wang and Brendan Quinn and Brian Fioca and Brian Hsu and Brian Yang and Brian Yu and Brian Zhang and Brittany Brenner and Callie Riggins Zetino and Cameron Raymond and Camillo Lugaresi and Carolina Paz and Cary Hudson and Cedric Whitney and Chak Li and Charles Chen and Charlotte Cole and Chelsea Voss and Chen Ding and Chen Shen and Chengdu Huang and Chris Colby and Chris Hallacy and Chris Koch and Chris Lu and Christina Kaplan and Christina Kim and CJ Minott-Henriques and Cliff Frey and Cody Yu and Coley Czarnecki and Colin Reid and Colin Wei and Cory Decareaux and Cristina Scheau and Cyril Zhang and Cyrus Forbes and Da Tang and Dakota Goldberg and Dan Roberts and Dana Palmie and Daniel Kappler and Daniel Levine and Daniel Wright and Dave Leo and David Lin and David Robinson and Declan Grabb and Derek Chen and Derek Lim and Derek Salama and Dibya Bhattacharjee and Dimitris Tsipras and Dinghua Li and Dingli Yu and DJ Strouse and Drew Williams and Dylan Hunn and Ed Bayes and Edwin Arbus and Ekin Akyurek and Elaine Ya Le and Elana Widmann and Eli Yani and Elizabeth Proehl and Enis Sert and Enoch Cheung and Eri Schwartz and Eric Han and Eric Jiang and Eric Mitchell and Eric Sigler and Eric Wallace and Erik Ritter and Erin Kavanaugh and Evan Mays and Evgenii Nikishin and Fangyuan Li and Felipe Petroski Such and Filipe de Avila Belbute Peres and Filippo Raso and Florent Bekerman and Foivos Tsimpourlas and Fotis Chantzis and Francis Song and Francis Zhang and Gaby Raila and Garrett McGrath and Gary Briggs and Gary Yang and Giambattista Parascandolo and Gildas Chabot and Grace Kim and Grace Zhao and Gregory Valiant and Guillaume Leclerc and Hadi Salman and Hanson Wang and Hao Sheng and Haoming Jiang and Haoyu Wang and Haozhun Jin and Harshit Sikchi and Heather Schmidt and Henry Aspegren and Honglin Chen and Huida Qiu and Hunter Lightman and Ian Covert and Ian Kivlichan and Ian Silber and Ian Sohl and Ibrahim Hammoud and Ignasi Clavera and Ikai Lan and Ilge Akkaya and Ilya Kostrikov and Irina Kofman and Isak Etinger and Ishaan Singal and Jackie Hehir and Jacob Huh and Jacqueline Pan and Jake Wilczynski and Jakub Pachocki and James Lee and James Quinn and Jamie Kiros and Janvi Kalra and Jasmyn Samaroo and Jason Wang and Jason Wolfe and Jay Chen and Jay Wang and Jean Harb and Jeffrey Han and Jeffrey Wang and Jennifer Zhao and Jeremy Chen and Jerene Yang and Jerry Tworek and Jesse Chand and Jessica Landon and Jessica Liang and Ji Lin and Jiancheng Liu and Jianfeng Wang and Jie Tang and Jihan Yin and Joanne Jang and Joel Morris and Joey Flynn and Johannes Ferstad and Johannes Heidecke and John Fishbein and John Hallman and Jonah Grant and Jonathan Chien and Jonathan Gordon and Jongsoo Park and Jordan Liss and Jos Kraaijeveld and Joseph Guay and Joseph Mo and Josh Lawson and Josh McGrath and Joshua Vendrow and Joy Jiao and Julian Lee and Julie Steele and Julie Wang and Junhua Mao and Kai Chen and Kai Hayashi and Kai Xiao and Kamyar Salahi and Kan Wu and Karan Sekhri and Karan Sharma and Karan Singhal and Karen Li and Kenny Nguyen and Keren Gu-Lemberg and Kevin King and Kevin Liu and Kevin Stone and Kevin Yu and Kristen Ying and Kristian Georgiev and Kristie Lim and Kushal Tirumala and Kyle Miller and Lama Ahmad and Larry Lv and Laura Clare and Laurance Fauconnet and Lauren Itow and Lauren Yang and Laurentia Romaniuk and Leah Anise and Lee Byron and Leher Pathak and Leon Maksin and Leyan Lo and Leyton Ho and Li Jing and Liang Wu and Liang Xiong and Lien Mamitsuka and Lin Yang and Lindsay McCallum and Lindsey Held and Liz Bourgeois and Logan Engstrom and Lorenz Kuhn and Louis Feuvrier and Lu Zhang and Lucas Switzer and Lukas Kondraciuk and Lukasz Kaiser and Manas Joglekar and Mandeep Singh and Mandip Shah and Manuka Stratta and Marcus Williams and Mark Chen and Mark Sun and Marselus Cayton and Martin Li and Marvin Zhang and Marwan Aljubeh and Matt Nichols and Matthew Haines and Max Schwarzer and Mayank Gupta and Meghan Shah and Melody Huang and Meng Dong and Mengqing Wang and Mia Glaese and Micah Carroll and Michael Lampe and Michael Malek and Michael Sharman and Michael Zhang and Michele Wang and Michelle Pokrass and Mihai Florian and Mikhail Pavlov and Miles Wang and Ming Chen and Mingxuan Wang and Minnia Feng and Mo Bavarian and Molly Lin and Moose Abdool and Mostafa Rohaninejad and Nacho Soto and Natalie Staudacher and Natan LaFontaine and Nathan Marwell and Nelson Liu and Nick Preston and Nick Turley and Nicklas Ansman and Nicole Blades and Nikil Pancha and Nikita Mikhaylin and Niko Felix and Nikunj Handa and Nishant Rai and Nitish Keskar and Noam Brown and Ofir Nachum and Oleg Boiko and Oleg Murk and Olivia Watkins and Oona Gleeson and Pamela Mishkin and Patryk Lesiewicz and Paul Baltescu and Pavel Belov and Peter Zhokhov and Philip Pronin and Phillip Guo and Phoebe Thacker and Qi Liu and Qiming Yuan and Qinghua Liu and Rachel Dias and Rachel Puckett and Rahul Arora and Ravi Teja Mullapudi and Raz Gaon and Reah Miyara and Rennie Song and Rishabh Aggarwal and RJ Marsan and Robel Yemiru and Robert Xiong and Rohan Kshirsagar and Rohan Nuttall and Roman Tsiupa and Ronen Eldan and Rose Wang and Roshan James and Roy Ziv and Rui Shu and Ruslan Nigmatullin and Saachi Jain and Saam Talaie and Sam Altman and Sam Arnesen and Sam Toizer and Sam Toyer and Samuel Miserendino and Sandhini Agarwal and Sarah Yoo and Savannah Heon and Scott Ethersmith and Sean Grove and Sean Taylor and Sebastien Bubeck and Sever Banesiu and Shaokyi Amdo and Shengjia Zhao and Sherwin Wu and Shibani Santurkar and Shiyu Zhao and Shraman Ray Chaudhuri and Shreyas Krishnaswamy and Shuaiqi and Xia and Shuyang Cheng and Shyamal Anadkat and Simón Posada Fishman and Simon Tobin and Siyuan Fu and Somay Jain and Song Mei and Sonya Egoian and Spencer Kim and Spug Golden and SQ Mah and Steph Lin and Stephen Imm and Steve Sharpe and Steve Yadlowsky and Sulman Choudhry and Sungwon Eum and Suvansh Sanjeev and Tabarak Khan and Tal Stramer and Tao Wang and Tao Xin and Tarun Gogineni and Taya Christianson and Ted Sanders and Tejal Patwardhan and Thomas Degry and Thomas Shadwell and Tianfu Fu and Tianshi Gao and Timur Garipov and Tina Sriskandarajah and Toki Sherbakov and Tomer Kaftan and Tomo Hiratsuka and Tongzhou Wang and Tony Song and Tony Zhao and Troy Peterson and Val Kharitonov and Victoria Chernova and Vineet Kosaraju and Vishal Kuo and Vitchyr Pong and Vivek Verma and Vlad Petrov and Wanning Jiang and Weixing Zhang and Wenda Zhou and Wenlei Xie and Wenting Zhan and Wes McCabe and Will DePue and Will Ellsworth and Wulfie Bain and Wyatt Thompson and Xiangning Chen and Xiangyu Qi and Xin Xiang and Xinwei Shi and Yann Dubois and Yaodong Yu and Yara Khakbaz and Yifan Wu and Yilei Qian and Yin Tat Lee and Yinbo Chen and Yizhen Zhang and Yizhong Xiong and Yonglong Tian and Young Cha and Yu Bai and Yu Yang and Yuan Yuan and Yuanzhi Li and Yufeng Zhang and Yuguang Yang and Yujia Jin and Yun Jiang and Yunyun Wang and Yushi Wang and Yutian Liu and Zach Stubenvoll and Zehao Dou and Zheng Wu and Zhigang Wang},
      year={2025},
      eprint={2601.03267},
      archivePrefix={arXiv},
      primaryClass={cs.CL},
      url={https://arxiv.org/abs/2601.03267}, 
}

@misc{dhar2025evalcardsframeworkstandardizedevaluation,
      title={EvalCards: A Framework for Standardized Evaluation Reporting}, 
      author={Ruchira Dhar and Danae Sanchez Villegas and Antonia Karamolegkou and Alice Schiavone and Yifei Yuan and Xinyi Chen and Jiaang Li and Stella Frank and Laura De Grazia and Monorama Swain and Stephanie Brandl and Daniel Hershcovich and Anders Søgaard and Desmond Elliott},
      year={2025},
      eprint={2511.21695},
      archivePrefix={arXiv},
      primaryClass={cs.CL},
      url={https://arxiv.org/abs/2511.21695}, 
}

@incollection{HENDERSON1995793,
title = {There's No Place Like Home: Continuing Design in Use},
editor = {RONALD M. BAECKER and JONATHAN GRUDIN and WILLIAM A.S. BUXTON and SAUL GREENBERG},
booktitle = {Readings in Human–Computer Interaction},
publisher = {Morgan Kaufmann},
pages = {793-803},
year = {1995},
series = {Interactive Technologies},
isbn = {978-0-08-051574-8},
doi = {https://doi.org/10.1016/B978-0-08-051574-8.50082-0},
url = {https://www.sciencedirect.com/science/article/pii/B9780080515748500820},
author = {Austin Henderson and Morten Kyng}
}

@misc{hamna2025buildingbenchmarksgroundup,
      title={Building Benchmarks from the Ground Up: Community-Centered Evaluation of LLMs in Healthcare Chatbot Settings}, 
      author={Hamna and Gayatri Bhat and Sourabrata Mukherjee and Faisal Lalani and Evan Hadfield and Divya Siddarth and Kalika Bali and Sunayana Sitaram},
      year={2025},
      eprint={2509.24506},
      archivePrefix={arXiv},
      primaryClass={cs.CL},
      url={https://arxiv.org/abs/2509.24506}, 
}

@misc{shrivastava2025diceframeworkdimensionalcontextual,
      title={DICE: A Framework for Dimensional and Contextual Evaluation of Language Models}, 
      author={Aryan Shrivastava and Paula Akemi Aoyagui},
      year={2025},
      eprint={2504.10359},
      archivePrefix={arXiv},
      primaryClass={cs.CL},
      url={https://arxiv.org/abs/2504.10359}, 
}

@article{sadek2023designing,
  author    = {Sadek, Malak and Calvo, Rafael A. and Mougenot, C{\'e}line},
  title     = {Designing value-sensitive AI: a critical review and recommendations for socio-technical design processes},
  journal   = {AI and Ethics},
  volume    = {4},
  number    = {4},
  pages     = {949--967},
  year      = {2023},
  doi       = {10.1007/s43681-023-00373-7},
  url       = {https://doi.org/10.1007/s43681-023-00373-7}
}

@article{10.1145/1629175.1629210,
author = {Khatri, Vijay and Brown, Carol V.},
title = {Designing data governance},
year = {2010},
issue_date = {January 2010},
publisher = {Association for Computing Machinery},
address = {New York, NY, USA},
volume = {53},
number = {1},
issn = {0001-0782},
url = {https://doi.org/10.1145/1629175.1629210},
doi = {10.1145/1629175.1629210},
journal = {Commun. ACM},
month = jan,
pages = {148–152},
numpages = {5}
}

@misc{russell2025aiuseamericannewspapers,
      title={AI use in American newspapers is widespread, uneven, and rarely disclosed}, 
      author={Jenna Russell and Marzena Karpinska and Destiny Akinode and Katherine Thai and Bradley Emi and Max Spero and Mohit Iyyer},
      year={2025},
      eprint={2510.18774},
      archivePrefix={arXiv},
      primaryClass={cs.CL},
      url={https://arxiv.org/abs/2510.18774}, 
}

@misc{chiang2024chatbotarenaopenplatform,
      title={Chatbot Arena: An Open Platform for Evaluating LLMs by Human Preference}, 
      author={Wei-Lin Chiang and Lianmin Zheng and Ying Sheng and Anastasios Nikolas Angelopoulos and Tianle Li and Dacheng Li and Hao Zhang and Banghua Zhu and Michael Jordan and Joseph E. Gonzalez and Ion Stoica},
      year={2024},
      eprint={2403.04132},
      archivePrefix={arXiv},
      primaryClass={cs.AI},
      url={https://arxiv.org/abs/2403.04132}, 
}

@misc{jiang2024genaiarenaopenevaluation,
      title={GenAI Arena: An Open Evaluation Platform for Generative Models}, 
      author={Dongfu Jiang and Max Ku and Tianle Li and Yuansheng Ni and Shizhuo Sun and Rongqi Fan and Wenhu Chen},
      year={2024},
      eprint={2406.04485},
      archivePrefix={arXiv},
      primaryClass={cs.AI},
      url={https://arxiv.org/abs/2406.04485}, 
}

@misc{wang2024factcheckbenchfinegrainedevaluationbenchmark,
      title={Factcheck-Bench: Fine-Grained Evaluation Benchmark for Automatic Fact-checkers}, 
      author={Yuxia Wang and Revanth Gangi Reddy and Zain Muhammad Mujahid and Arnav Arora and Aleksandr Rubashevskii and Jiahui Geng and Osama Mohammed Afzal and Liangming Pan and Nadav Borenstein and Aditya Pillai and Isabelle Augenstein and Iryna Gurevych and Preslav Nakov},
      year={2024},
      eprint={2311.09000},
      archivePrefix={arXiv},
      primaryClass={cs.CL},
      url={https://arxiv.org/abs/2311.09000}, 
}

@misc{rottger2025issuebenchmillionsrealisticprompts,
      title={IssueBench: Millions of Realistic Prompts for Measuring Issue Bias in LLM Writing Assistance}, 
      author={Paul Röttger and Musashi Hinck and Valentin Hofmann and Kobi Hackenburg and Valentina Pyatkin and Faeze Brahman and Dirk Hovy},
      year={2025},
      eprint={2502.08395},
      archivePrefix={arXiv},
      primaryClass={cs.CL},
      url={https://arxiv.org/abs/2502.08395}, 
}

@techreport{bbcebu2025newsintegrity,
  author      = {{BBC} and {European Broadcasting Union}},
  title       = {News Integrity in {AI} Assistants: An International {PSM} Study},
  institution = {BBC and European Broadcasting Union (EBU)},
  year        = {2025},
  month       = {October},
  url         = {https://www.ebu.ch/files/live/sites/ebu/files/Publications/MIS/open/EBU-MIS-BBC_News_Integrity_in_AI_Assistants_Report_2025.pdf}
}

@inproceedings{10.1145/3613904.3642278,
author = {Kuo, Tzu-Sheng and Halfaker, Aaron Lee and Cheng, Zirui and Kim, Jiwoo and Wu, Meng-Hsin and Wu, Tongshuang and Holstein, Kenneth and Zhu, Haiyi},
title = {Wikibench: Community-Driven Data Curation for AI Evaluation on Wikipedia},
year = {2024},
isbn = {9798400703300},
publisher = {Association for Computing Machinery},
address = {New York, NY, USA},
url = {https://doi.org/10.1145/3613904.3642278},
doi = {10.1145/3613904.3642278},
booktitle = {Proceedings of the 2024 CHI Conference on Human Factors in Computing Systems},
articleno = {193},
numpages = {24},
location = {Honolulu, HI, USA},
series = {CHI '24}
}

\appendix
\clearpage
\section{Rapporteur Discussion Template}
\label{appendix:b}
\subsection{Use-Case Breakout Group}
\begin{itemize}
    \item \textbf{Part 1: How have you used or would use generative AI for this use case?} Write down the content/context for which gen AI could be used. What task would be given as input to the system? What is a typical output?
    \begin{itemize}
        \item \textit{Discussion Question}: Of these cases, which are more important to benchmark, in terms of prevalence, journalistic impact, audience relevance, organizational context and priorities, etc? 
    \end{itemize}
    \item \textbf{Part 2: How to assess the performance on specific tasks in this use case?} For the more important tasks selected in the discussion earlier, write down elements of a high-quality outcome for this use case. Who gets to determine this? What are some pitfalls one might encounter?
    \begin{itemize}
        \item \textit{Discussion Question}: How might these elements be summarized as general criteria for success?
    \end{itemize}
    \item \textbf{Part 3: How would you set up a test case to evaluate a generative AI system’s performance for the specific tasks discussed?} How might you prompt the system (e.g. inputs and/or context) in such a test? What would be considered a “good” output vs a “bad” output? How does the system “correctly” respond to the input?
    \begin{itemize}
        \item \textit{Discussion Question}: How do you ensure the prompt and answer pairs are measuring what you intend for them to measure in a system?
    \end{itemize}
    \item \textbf{Part 4: Are there any existing datasets that can be used to create these test cases? What data would we need to collect (and from whom) to make such test cases?}
    \begin{itemize}
        \item \textit{Discussion Question}: How well would these datasets align with the way the use case is really practiced? What are challenges or issues we might encounter using or collecting these datasets? 
    \end{itemize}
\end{itemize}

\subsection{Value Breakout Group}
\begin{itemize}
    \item \textbf{Part 1: How is this value reflected in your work?} Write down specific cases where this value is applied in journalistic work? What are some tensions and barriers in complying with value?
    \begin{itemize}
        \item \textit{Discussion Question}: How would you define this value in journalism?
    \end{itemize}
    \item \textbf{Part 2: How do you imagine a generative AI system could violate this value?} Write down scenarios where a generative AI system could violate this value or demonstrate that it is aligned with the value.
    \begin{itemize}
        \item \textit{Discussion Question}: Based on the examples, how would you define when a LM aligns with this value?
    \end{itemize}
    \item \textbf{Part 3: How would you set up a test case to assess the alignment of a generative AI model or system with this value?} How might you prompt the system (e.g. inputs and/or context) in such a test? What would be considered a “good” output vs a “bad” output? How does the system “correctly” respond to the input? 
    \begin{itemize}
        \item \textit{Discussion Question}:How do you ensure the prompt and answer pairs are measuring what you intend for them to measure in a system?
    \end{itemize}
    \item \textbf{Part 4: Are there any existing datasets that can be used to create these test cases? What data would we need to collect (and from whom) to make such test cases?}
    \begin{itemize}
        \item \textit{Discussion Question}: How well would these datasets align with the way the use case is really practiced? What are challenges or issues we might encounter using or collecting these datasets? 
    \end{itemize}
\end{itemize}

\clearpage
\onecolumn
\section{Workshop Structure}
\label{appendix:a}
\newcolumntype{L}{>{\RaggedRight\hangafter=1\hangindent=1em}X}
\begin{table*}[h]
    \centering
    \begin{tabularx}{\textwidth}{@{} l l l L @{}}
    \toprule
    \textbf{Session (shorthand)} & \textbf{Topic} & \textbf{Participants} & \textbf{Example in Journalism}  \\
    \toprule
     Introduction (\textbf{INT}) & Intro to Existing Practices & P1-23 &\\
     \hline
     Use Case Breakout 1 (\textbf{UC1}) & Information/Data Extraction &  P3, P4, P7, P13 & Extracting entities from documents \\ 
     Use Case Breakout 2 (\textbf{UC2}) & Semantic Search & P11, P14, P17, P22 & Searching through transcripts for concepts  \\
     Use Case Breakout 3 (\textbf{UC3}) & Summarization &  P6, P10, P12, P23 & Creating bullet point summaries of news stories to be included at the top of an article page \\
     Use Case Breakout 4 (\textbf{UC4}) & Content Transformation &  P9, P20, P21 & Converting a news story into a set of quiz questions \\
     Use Case Breakout 5 (\textbf{UC5}) & Background Research & P2, P8, P15, P16 & Identifying relevant documents from an archive for a news topic  \\
     Use Case Breakout 6 (\textbf{UC6}) & Fact Checking & P1, P5, P18, P19 & Verifying the accuracy of claims in a politician's speech \\
     \hline
      Value Breakout 1 (\textbf{V1}) & Accuracy &  P13, P16, P19, P21 &Reflects a commitment to seeking the truth \\
     Value Breakout 2 (\textbf{V2}) & Transparency &  P1, P9, P11, P15 &Disclosure of information about practices, methods, and sources \\
     Value Breakout 3 (\textbf{V3}) & Confidence/Uncertainty & P20, P22, P23 & Degree of trust or skepticism about claims given available evidence \\
     Value Breakout 4 (\textbf{V4}) & Accountability & P2, P6, P7, P18 &Responsibility to the public for what is published and adherance to standards and ethics \\
     Value Breakout 5 (\textbf{V5}) & Objectivity/Bias & P4, P5, P8, P12 & Rigorous, open-minded reporting methods that mitigate bias \\
     Value Breakout 6 (\textbf{V6}) & Timeliness/Recency & P3, P10, P14, P17 & Reflecting a commitment to immediacy, relevance in time, and freshness \\
    \hline
    Synthesis (\textbf{SYN}) & Challenges, Takeaways, Next Steps & P1-23 & \\
    \\
    \end{tabularx}
    \caption{Sessions conducted during the workshop listed in the order in which they happened. The examples for use cases draw on related work on the use of generative AI in journalism such as \cite{cools2024uses-635, 10.13140/RG.2.2.31540.05765} and value breakout session examples draw from research such as  \cite{10.1145/3715336.3735717, diakopoulos2023leveraging-af4, Deuze2005if, Hanitzsch2007it}. We use the session shorthand in the body of the paper when describing findings grounded in specific sessions.}.
    \label{tab:sessions}
    \Description{Table with 4 columns and 4 groups of rows. Column-wise, they indicate session shorthand, topic discussed during the session, participants in the session represented by their participant number, and an example in journalism related to the topic discussed in the session. Row-wise, rows are divided into 4 groups representing the timing of sessions in order of introduction, use case (several happening simultaneously), value (several happening simultaneously), and synthesis at the end.}
\end{table*}

\end{document}